\documentclass[aps,pra,epsf,preprint,superscriptaddress,showpacs,preprintnumbers,amsmath,amssymb,amsfonts,onecolumn] {revtex4-1}
\usepackage[utf8]{inputenc}
\usepackage{enumitem}
\usepackage{amsfonts}
\usepackage{amsmath}
\usepackage{amssymb}
\usepackage{dsfont}
\usepackage{color}
\usepackage{graphicx}
\usepackage{textcomp}
\usepackage{cancel}
\usepackage{dcolumn}
\usepackage{bm}
\usepackage{braket}
\usepackage{float}
\usepackage{mathrsfs} 
\usepackage{todonotes}
\def\be{\begin{equation}}
\def\ee{\end{equation}}
\def\bea{\begin{eqnarray}}
\def\eea{\end{eqnarray}}
\newcommand{\bs}{\begin{split}}
\newcommand{\es}{\end{split}}                            
\newcommand{\bear}{\begin{eqnarray}}
\newcommand{\eear}{\end{eqnarray}}

\newcommand{\abs}[1]{\left| #1 \right|}

\newcommand   {\<}           { \langle }
\renewcommand {\>}           { \rangle }
\renewcommand{\bra}[1]{\< #1|}	
\renewcommand{\ket}[1]{|#1\>}

\newcommand{\vecB} [1] {\mb{#1} }	
\renewcommand{\vec} [1] { {\vecB{#1}} } 

\newcommand{\mb}[1]{\bm{#1}}	

\newcommand{\Hut}[1]{\hat{#1}}		
\newcommand{\wdag} { {\textcolor{white}{\dagger}} }   
    \definecolor{background}{gray}{0.8}

\newcommand{\psiSb}  [2][\sigma] { \bra{\psi_{#2}^{#1}} }
\newcommand{\psiSk}  [2][\sigma] { \ket{\psi_{#2}^{#1}} }

\newcommand{\psiSkt}  [2][\sigma] { \ket{\psi_{#2}^{#1}(t)} }

\newcommand{\psiSlabel}  [2][\sigma] {\psi_{#2}^{#1}}
\newcommand{\nb}  [2][\sigma] { {}^{#1}\bra{\vec{#2}} }
\newcommand{\nk}  [2][\sigma] { \ket{\vec{#2}}^{#1} }

\newcommand{\phiPb}  [2][\sigma] { \bra{\phi_{#2}^{#1}} }
\newcommand{\phiPk}  [2][\sigma] { \ket{\phi_{#2}^{#1}} }

\newcommand{\phiPkt}  [2][\sigma] { \ket{\phi_{#2}^{#1}(t)} }
\newcommand{\phiSlabel}  [2][\sigma] {\phi_{#2}^{#1}}
\newcommand{\phiPsk}  [3][\sigma] { {\phi_{#3}^{#1}}(#2) }
\newcommand{\phiPsb}  [3][\sigma] { {\big(\phi_{#3}^{#1}}(#2)\big)^*}

\newcommand{\ad}    [2][\sigma] { \Hut{a} ^\dagger _{ #2,#1 } }
\renewcommand{\a}   [2][\sigma] { \Hut{a} ^\wdag   _{ #2,#1 } }

\begin{document}
\title{A unified ab-initio approach to the correlated quantum dynamics\\ of ultracold fermionic and bosonic 
mixtures}

\author{L. Cao}
\email{lushuai\_cao@hust.edu.cn}
\affiliation{Zentrum f\"ur Optische Quantentechnologien, Universit\"at Hamburg, Luruper Chaussee 149,
22761 Hamburg, Germany}
\affiliation{Ministry of Education Key Laboratory of Fundamental Physical Quantities Measurements, School of 
Physics, Huazhong University of Science and Technology, Wuhan 430074, People’s Republic of China}
\author{ V. Bolsinger}
\affiliation{Zentrum f\"ur Optische Quantentechnologien, Universit\"at Hamburg, Luruper Chaussee 149,
22761 Hamburg, Germany}
\affiliation{The Hamburg Centre for Ultrafast Imaging, Universit\"at Hamburg, Luruper Chaussee 149, 22761
Hamburg, Germany}
\author{S. I. Mistakidis}
\affiliation{Zentrum f\"ur Optische Quantentechnologien, Universit\"at Hamburg, Luruper Chaussee 149,
22761 Hamburg, Germany} 
\author{ G. M. Koutentakis}
\affiliation{Zentrum f\"ur Optische Quantentechnologien, Universit\"at Hamburg, Luruper Chaussee 149,
22761 Hamburg, Germany}
\affiliation{The Hamburg Centre for Ultrafast Imaging, Universit\"at Hamburg, Luruper Chaussee 149, 22761
Hamburg, Germany}
\author{S. Kr\"onke}
\affiliation{Zentrum f\"ur Optische Quantentechnologien, Universit\"at Hamburg, Luruper Chaussee 149,
22761 Hamburg, Germany}
\affiliation{The Hamburg Centre for Ultrafast Imaging, Universit\"at Hamburg, Luruper Chaussee 149, 22761
Hamburg, Germany}
\author{J. M. Schurer}
\affiliation{Zentrum f\"ur Optische Quantentechnologien, Universit\"at Hamburg, Luruper Chaussee 149,
22761 Hamburg, Germany}
\affiliation{The Hamburg Centre for Ultrafast Imaging, Universit\"at Hamburg, Luruper Chaussee 149, 22761
Hamburg, Germany}
\author{P. Schmelcher}
\email{pschmelc@physnet.uni-hamburg.de}
\affiliation{Zentrum f\"ur Optische Quantentechnologien, Universit\"at Hamburg, Luruper Chaussee 149,
22761 Hamburg, Germany}
\affiliation{The Hamburg Centre for Ultrafast Imaging, Universit\"at Hamburg, Luruper Chaussee 149, 22761
Hamburg, Germany}

\begin{abstract}
We extent the recently developed Multi-Layer Multi-Configuration Time-Dependent Hartree method for Bosons 
(ML-MCTDHB) for simulating the correlated quantum dynamics of bosonic mixtures to the fermionic sector and establish a 
unifying approach for the investigation of the correlated quantum dynamics of mixture of indistinguishable 
particles,
be it fermions or bosons. Relying on a multi-layer wave-function expansion, the resulting
Multi-Layer Multi-Configuration Time-Dependent Hartree method for Mixtures (ML-MCTDHX) can be adapted to 
efficiently resolve system-specific intra- and inter-species correlations. The versatility and efficiency of
ML-MCTDHX is demonstrated by applying it to the problem of colliding few-atom mixtures of both Bose-Fermi and 
Fermi-Fermi type. Thereby, we elucidate the role of correlations in the transmission and reflection 
properties of the collisional events. In particular, we present examples where the reflection (transmission)
at  the other atomic species is a correlation-dominated effect, i.e.\ it is suppressed in the mean-field 
approximation.
\end{abstract}

\maketitle


\section{Introduction}

Degenerate Bose and Fermi gases allow for an exquisite experimental control of the inter-atomic interactions 
as well as the motional and the internal quantum states which renders them ideal prototypes to investigate  
many-body quantum phenomena~\cite{Bloch2008c}. This holds in particular for mixtures of different components
which could be different chemical species or isotopes as well as different spin degrees of freedom 
represented by the hyperfine states. Depending on the statistics of the species, one 
can realize Bose-Bose (BB)~\cite{bbmix}, Fermi-Fermi (FF)~\cite{ff0,ff1} and Bose-Fermi (BF) 
mixtures~\cite{bf0,bf1,bf2} with or without pseudo-spin degrees of freedom. 

These ultracold mixtures reveal a great variety of physical phenomena depending among others on the particle 
statistics, the interaction strength and the external confinement. Examples for phenomena in BB systems are 
the relative phase  evolution~\cite{Anderson2009}, composite fermionization and phase 
separation~\cite{Garcia-March2013,Zollner2008a}, binary non-linear~\cite{Kamchatnov2013} or 
collective~\cite{Mertes2007} excitations, and the miscible to immiscible phase 
transition~\cite{Ticknor2013,Nicklas2015}.
For the corresponding spinor systems~\cite{Kawaguchi2012,Stamper-Kurn2013} in optical lattices, their 
ferromagnetic and antiferromagnetic behavior~\cite{Eisenberg2002,Sadler2006}, the interplay of spin and 
charge excitations~\cite{Zvonarev2007}, governed  by spin-charge 
separation in one dimension~\cite{Kleine2008}, and spin-exchange interactions~\cite{Lee2009} have been investigated.\\
\noindent
In nature, material properties are often determined by the underlying electronic structure and thus in 
particular fermionic degrees of freedom play a decisive role. By 
advanced cooling techniques~\cite{Onofrio}, ultracold fermionic gases can nowadays be prepared with deterministic particle number and quantum state 
control~\cite{Serwane2011} even for two-particle systems~\cite{Murmann2015} freeing the way to observe 
the onset of many-body physics by successively increasing the particle number~\cite{Wenz2013}. These systems 
exhibit fermion pairing~\cite{Zurn2013} and Wigner molecule formation~\cite{Brandt2015} 
and are ideally suited to investigate the physics of 
magnetism~\cite{Murmann2015a,Yannouleas2016,Deuretzbacher2014,Volosniev2014a}, spin dynamics~\cite{Koller2015,Volosniev2015,Volosniev2016}, and  
collective excitations~\cite{Koller2016}. For two species mixtures of fermions a transition from a molecular 
bosonic condensate~\cite{Strecker2003} to a Bardeen-Cooper-Schrieffer super-fluid of weakly bound 
Cooper pairs~\cite{Stoof1994,Zwierlein2004,Ries2015} has been observed. For strong inter-species interaction, 
even such distinguishable fermions can be fermionized~\cite{Zurn2012}. \\
\noindent
In the same spirit, binary BF mixtures can be mapped to a FF system for strong inter-bosonic
interactions~\cite{Girardeau2007} being describable by the Gaudin-Yang model~\cite{Gaudin1967, Yang1967}.
The low energy physics can be described by the Tomonaga-Luttinger liquid theory~\cite{Cazalilla2003} 
revealing a charge-density wave phase, a fermion pairing phase, and a phase separation 
regime~\cite{Mathey2004} comparable to a liquid of polarons. Moreover, collective 
oscillations such as the monopole~\cite{Maruyama2005}, dipole~\cite{Maruyama2006}, and 
scissors~\cite{Roy2017} mode have been analyzed for BF mixtures. For attractive interactions, collapse~\cite{Ospelkaus2006} or 
the formation of composite particles, i.e.\ bound boson fermion pairs, becomes 
possible~\cite{Varney2008,Burovski2009}. Additionally, these systems are ideally suited to understand 
so-called induced interactions~\cite{Kinnunen2015} and their impact on the super-fluid transition 
temperature~\cite{Heiselberg2000}. 

In order to capture the above indicated variety of quantum phenomena, one needs a highly flexible theoretical 
description which is able to go beyond mean-field approximations such as the Gross-Pitaevskii 
(GP)~\cite{Pitaevskii61,gross_hydrodynamics_1963} or the Hartree-Fock 
(HF)~\cite{hartree_wave_1928,fock_naherungsmethode_1930} theory, and includes inter- and intra- species 
correlations.
Since analytically solvable many-body problems are rare 
\cite{exact_analysis_interacting_Bose_gas_I_Lieb_Liniger,exact_analysis_interacting_Bose_gas_II_Lieb_Liniger,
2bosons_in_harmonic_trap_Busch,Klaiman2017362} 
and exact diagonalization is restricted to very few degrees of freedom, sophisticated methods to solve 
the time-dependent many-body Schr\"odinger equation of mixture systems from first principles are required.
Various wave-function propagation methods face the exponential scaling of complexity with the number
of particles by expanding the system state $|\Psi_t\rangle$ w.r.t.\ a dynamically optimized truncated basis. 
This family of methods may be divided into two classes with different underlying concepts
of correlations. On the one hand, there are methods, in which correlations are regarded
as deviations from a state that factorized w.r.t.\ to certain {\it a-priori} given
modes, i.e.\ from a Gutzwiller mean-field state. Famous examples are the tensor network
methods based on matrix-product or projected entangled-pair states
\cite{schollwock_density-matrix_2005,MPS_PEPs_rev08,DMRG_at_age_of_MPS_Schollwoeck2010},
which are mostly tailored to lattice problems and proved to be highly successful in particular
for one-dimensional systems.
On the other hand, correlations can be defined as deviations from
a, possibly (anti-) symmetrized, product state w.r.t.\ particles. Here, the correlation-free
reference state is a single
Hartree product in the case of distinguishable particles or indistinguishable bosons and 
a single Slater determinant in the case of fermions, respectively. This is exactly the
correlation concept underlying the Multi-Configuration Time-Dependent
Hartree (MCTDH) family of methods where the total state 
$|\Psi_t\rangle$ is expanded w.r.t.\ a time-dependent many-body basis. Originally designed for 
distinguishable 
degrees of freedom \cite{mctdh_approach_Meyer_Manthe_Cederbaum,MCTDH_BJMW2000}, MCTDH became 
tailored to indistinguishable fermions and bosons in terms of the
MCTDHF \cite{MCTDHF_approach_to_multi_electr_dynamics_in_laser_fields_Scrinzi2003,kato_time-dependent_2004,
nest_multiconfiguration_2005} 
and MCTDHB method 
\cite{role_of_excited_states_in_splitting_of_BEC_by_time_dep_barrier_Steltsov_PRL2007,Alon2008}, by using 
Slater determinants and permanents of variationally
optimized time-dependent single-particle states as a many-body basis, respectively, which can 
formally be treated in a unifying manner 
\cite{unified_view_on_mactdh_for_identical_particles_Alon_Streltsov_Cederbaum_JChemPhys07,MCTDHX_recursively}.
Recent developments cover the usage of Wannier basis states for lattice systems 
\cite{Cao2015,lode_dynamics_2016},
internal degrees of freedom \cite{lode_multiconfigurational_2016}, linear-response theory 
\cite{exc_spectra_of_fragmented_condensates_by_lin_response_Grond_PRA2012,alon_unified_2014} and
restricting the active configuration space \cite{miyagi_time-dependent_2013,leveque_time-dependent_2016}.
A direct extension of MCTDHB and MCTDHF to bosonic, fermionic or BF mixtures 
\cite{mctdh_for_mixtures_of_2types_of_identical_particles_PRA2007,MCTDHX_recursively}, however,
is impractical except for very small system sizes due to the exponential wall of complexity.
To pursue, the concept of the so-called Multi-Layer MCTDH method 
\cite{Multilayer_formulation_Wang_Thoss_2003,Multilayer_multiconfig_Manthe_2008,
ML_implementation_application_Henon_Heiles_Pyrazine_Vendrell_Meyer_JChemPhys_2011}, 
namely the coarse-graining of degrees of freedom, has been applied to bosonic mixtures
in the recently developed Multi-Layer MCTDH method for Bosons (ML-MCTDHB) 
\cite{kronke_non-equilibrium_2013,cao_multi-layer_2013}. 
Here the central idea is to expand the state of the bosonic mixtures w.r.t.\
dynamically optimized truncated basis states for the constituting species, which are
in turn efficiently represented as MCTDHB wave functions. Thereby, 
the variational wave-function ansatz can be tailored to system-specific intra- and inter-species correlations.
In passing, we note that the multi-layering idea has also been applied directly to the structure of Fock 
space in the so-called Multi-Layer MCTDH method in (optimized) Second Quantization Representation
\cite{MLMCTDH_in_2nd_quantization_representation_Wang_Thoss_Chem_Phys_2009,ml_oSQR_Manthe17}, which is based 
on the concept of correlations between (optimized) modes.

The above variational approaches have already been applied successfully to ultracold systems. Beginning from 
one-dimensional single species systems, correlated processes in e.g.\ the 
tunneling~\cite{Zollner2008c,Sakmann2009,lode_how_2012,Schurer2016}, breathing~\cite{Schmitz2013}, 
soliton~\cite{Streltsov2011,Kronke2015b}, and quench induced 
dynamics in lattice systems~\cite{Mistakidis2015,beinke_many-body_2016,Koutentakis2017} including periodic 
driving~\cite{Mistakidis2017}, hybrid atom-ion systems~\cite{Schurer2015}, and 
dipolar systems~\cite{Cao2017,Streltsov_long_range,Budhaditya} have been explored.
In the realm of binary mixture systems, energy transfer~\cite{Kronke2015c}, mixture 
tunneling~\cite{Pflanzer2009,impact_of_spatial_correl_on_tunneling_mixture_Capo2012}, mesoscopic charged molecules~\cite{Schurer2017}, 
and dark-bright solitons~\cite{Katsimiga} have been investigated. Meanwhile, 
MCTDH type methods have also been applied to two- and three-dimensional ultracold atomic ensembles~\cite{Klaiman2014,Cao2015,Bolsinger2017,Wells2015,Weiner2017} for analyzing e.g.\ the dimensional
cross-over or vortex dynamics.

In this work, we pursue the path by the ML-MCTDHB approach further in order to derive and apply an 
efficient method for simulating the correlated quantum dynamics of {\it arbitrary} ultracold atomic mixtures,
let them be of bosonic, fermionic or Bose-Fermi type. For this purpose, the ML-MCTDHB wave-function
ansatz is generalized to also include fermionic species and a unifying approach is applied, which does not 
distinguish between the particle-statistics.
Using this ML-MCTDHX wave-function representation, we derive the equations of motion (EOM) for the expansion 
coefficients as well as the variational optimized basis states.
Hereupon, we apply the ML-MCTDHX method to the quantum dynamics of colliding small 
binary mixtures of BF and FF type in one spatial dimension demonstrating the applicability of our ML-MCTDHX 
approach. Collisions of ultracold atomic clouds, being investigated 
theoretically~\cite{Zin1,Ogren,Deuar,Chwedenczuk,Zin,Bach} as well as 
experimentally~\cite{Perrin,Bongs,Vogels,Kasevich}, can be strongly influenced by correlations 
becoming observable in the transmission and reflection properties~\cite{Deuar,Norrie,Norrie1}. We reveal such 
correlation-induced transmission and reflection in a BF and a FF mixture, respectively. Moreover, we 
show that analyzing the many-body wave-function in terms of the natural orbitals of the species and the 
one-body density matrices allows for dedicated insights into the beyond mean-field dynamics. 

This work is structured as follows. In section~\ref{chap:theory}, we derive the ML-MCTDHX method in 
detail. Starting with the wave-function ansatz, we compute and discuss the variational EOM in detail. 
Section~\ref{chap:app}  is devoted to the collisional dynamics of a BF and a FF mixture. 
After an exemplary detailed convergence study, we analyze the temporal evolution of the mixtures especially 
focusing on the impact of correlations. Finally, we provide our conclusion in section~\ref{chap:conc}.


\section{Theory}\label{chap:theory}
In this section, we explicate the working principles of the ML-MCTDHX method for
multi-component quantum gases by considering a binary mixture of
indistinguishable particles. The various possibilities of having a
BB, FF or BF mixture are treated within the same
formalism and the particle-statistics turns out to enter only in the evaluation of
certain entities. Discussing first the physical systems we aim to address, we then state
the ML-MCTDHX wave-function ansatz and derive the corresponding equations 
of motion. Thereafter, we comment on limiting cases, convergence and the 
generalization of the binary mixture case to systems with more than two species. Finally, it is briefly
addressed how we can treat atoms in more than one spatial dimensions and how we can incorporate 
internal degrees of freedom such as (pseudo-)spin into our
approach.

\subsection{Physical systems}
In order to demonstrate the main idea of ML-MCTDHX, we consider a mixture of two
species labeled with $A$ and $B$, respectively. Each species $\sigma=A,B$
shall be constituted by a fixed number of atoms $N_\sigma$ and
can be either fermionic or bosonic.
For simplicity, we first assume that the particles may only move in
one-dimensional space and that their internal degrees of freedom, if existing,
are frozen. The Hamiltonian of the binary mixture
\begin{equation}\label{eq:hamil}
 \hat{H} = \sum_{\sigma=A,B} \left[ \hat{H}_\sigma + \hat{V}_\sigma \right] +
\hat{W}_{AB},
\end{equation}
is decomposed into the single-particle and intra-species interaction terms of
the $\sigma$ species, $\hat{H}_\sigma$ and $\hat{V}_\sigma$, respectively, as
well as the inter-species interaction $\hat{W}_{AB}$. For being concrete, we
consider a model with atoms in continuous space (treating discrete
lattice systems is straightforward, see e.g.\ \cite{Cao2015,lode_dynamics_2016}), where the
single-particle Hamiltonian typically reads
\begin{equation}\label{eq:hamil_sing}
 \hat{H}_\sigma = \sum_{i=1}^{N_\sigma} \left[ \frac{ (\hat{p}_i^{\sigma})^2 }{2\mathcal{M}_\sigma}  +
u_\sigma(\hat{x}_i^\sigma) \right]\equiv\sum_{i=1}^{N_\sigma}\hat{h}^\sigma_i.
\end{equation}
Here $\mathcal{M}_\sigma$ denotes the mass of a $\sigma$-species particle and
$\hat{x}_i^\sigma$, $\hat{p}_i^\sigma$ refer to the position, momentum operator
of the $i$-th $\sigma$-species particle, respectively. $u_\sigma(\hat{x})$ is an arbitrary,
in general, species-selective external potential. In this paper, we are solely
concerned with two-body interactions \footnote{The extension to e.g.\ three-body terms
is conceptually straightforward}:
\begin{eqnarray}\label{eq:hamil_intra}
 \hat V_\sigma = \sum_{1\leq i<j \leq N_\sigma} v_\sigma(\hat x_i^\sigma,
		\hat x_j^\sigma),\\ \label{eq:hamil_inter}
 \hat W_{AB} = \sum_{i=1}^{N_A}\sum_{j=1}^{N_{B}}
		w_{AB}(\hat x_i^A,\hat x_j^B),
\end{eqnarray}
where in the context of ultracold atoms the interaction potentials
$v_\sigma(x_1,x_2)$, $w_{AB}(x_1,x_2)$ are usually of contact or dipolar
type \cite{Pitaevskii_Stringari_Bose-Einstein_Condensation2003,
Bose-Einstein_Condensation_Dilute_Gases_Pethick_Smith_2008}. As ML-MCTDHX constitutes a wave-function 
propagation method, all terms of the Hamiltonian may be time-dependent.

As commonly done, we discretize the model at hand by assigning a typically large 
but finite number of $G_\sigma$ single-particle basis states to each particle
of species $\sigma$, for instance a discrete-variable representation (DVR) 
\cite{dvr_and_their_utilization_Light_Carrington_2000,MCTDH_BJMW2000}
or, for lattice systems in the tight-binding approximation, Wannier states.
These states span the single-particle Hilbert space of the $\sigma$-species
$\mathfrak{h}^\sigma$. In the following, we assume that this discretization is fine enough to resolve the
relevant physics such that we may refer to
$\mathscr{H}=(\hat S_A\otimes_{i=1}^{N_A}\mathfrak{h}
^A_i) \bigotimes\,(\hat S_B\otimes_{j=1}^{N_B}\mathfrak{h} ^B_j)$
as the complete Hilbert space of the binary mixture. Here,
$\mathfrak{h}^\sigma_i\equiv \mathfrak{h}^\sigma$ and $\hat S_\sigma$
denotes the (anti-)symmetrization operator if the species $\sigma$ is bosonic
(fermionic). The complete Hilbert space 
$\mathscr{H}$ is typically far too high-dimensional for simulating quantum
dynamics. In order to face this problem, the state of the total system is
expanded with respect to a dynamically optimized, truncated many-body basis in
ML-MCTDHX, whose construction is the subject of the subsequent section.

\subsection{Construction of the many-body basis}\label{sec:wfn_ansatz}
In the ML-MCTDHX method, the state of the binary mixture $|\Psi(t)\rangle$ is 
firstly expanded as
\begin{equation}\label{eq:PSI}
    |\Psi(t)\rangle = \sum_{i,j=1}^{M} A_{ij}(t)\;   \psiSkt[A]{i} 
\psiSkt[B]{j}.
\end{equation}
In the above expansion, most importantly, not only 
the coefficients $A_{ij}$ are time-dependent but also the states
$\psiSkt{i}$ themselves.
The states $\{ \psiSkt{i}\}_{i=1}^M$ with $M\leq\dim{(\hat S_\sigma\otimes_{i=1}^{N_\sigma}\mathfrak{h}
^\sigma_i)}$ form a set of orthonormal functions for the $\sigma$ species, which can be 
viewed as a (truncated) orthonormal species basis \footnote{To avoid redundant terms, the same upper 
summation 
bound $M$ can be chosen for both species (see Sec.~\ref{sec_generalization}).}.
It is important to notice that these species basis states (SBSs) 
do not form a complete basis in the full 
Hilbert space of the $\sigma$ species $\hat S_\sigma\otimes_{i=1}^{N_\sigma}\mathfrak{h} 
^\sigma_i$ in general but only represent a basis in the time-dependent truncated sub-space being
relevant for the expansion \eqref{eq:PSI}. 

The SBSs are now constructed in the following way. First, we assign an orthonormal set
of $m_\sigma$ 
time-dependent single-particle functions (SPFs) $\{\phiPkt{n}\}_{n=1}^{m_\sigma}$ to each species, thereby
truncating the 
single-particle Hilbert space $\mathfrak{h}^\sigma$ to a lower-dimensional, time-dependent sub-space. 
Second, given this time-dependent truncated single-particle 
space ${\rm span}\{\phiPkt{n}\}_{n=1}^{m_\sigma}$, the correspondingly 
available truncated sub-space for the $N_\sigma$ particles of kind $\sigma$ is 
spanned by the occupation number states
\begin{equation}\label{eq:num_state}
 \nk{n}_t = 
\frac{1}{\sqrt{n_1!n_2!...n_{m_\sigma}!}}(\ad{1})^{n_1}(\ad{2})^{n_2}...(\ad{m_\sigma})^{n_{m_\sigma}}\ket{
vac 
},
\end{equation}
where $\ket{vac}$ is the vacuum state and $\ad{i}$ ($\a{i}$) creates \footnote{Note
that the order of creation operators is important 
for fermions and fixed by definition \eqref{eq:num_state}.}
(annihilates) one particle in state
$\phiPkt{i}$. The second quantization operators 
obey the (anti-)commutation relations
\begin{align}
 \left[ \a{k}, \ad[\sigma]{q} \right]_\pm &= \delta_{k,q},  \label{eq:anticom}\\
 \left[ \ad{k}, \ad[\sigma]{q} \right]_\pm &= \left[ \a{k}, \a[\sigma]{q}
\right]_\pm =0,
\end{align}
with $[\hat A,\hat B]_\pm = \hat A\hat B \pm \hat B\hat A$ where the $+$ and the $-$ correspond to the case 
of the $\sigma$ particles being fermions or bosons, respectively.
The vector $\vec{n} = (n_1,n_2,\cdots,n_{m_\sigma})$ with integers $n_r$ (for a fermionic species 
$n_r\in\{0,1\}$) determines the number of particles in a certain single-particle state with the restriction 
$\sum_{r=1}^{m_\sigma}n_r=N_\sigma$ such that $K_\sigma = \binom{N_\sigma+m_\sigma-1}{m_\sigma -1}$ 
[$K_\sigma = \binom{m_\sigma}{N_\sigma}$] different number states for $N_\sigma$ bosons (fermions) exist.

Finally, we truncate the available space ${\rm span}\{\nk{n}_t\}$ by considering only a
reduced number $M \leq \min\{K_A,K_B\}$ of optimally chosen time-dependent, pairwise orthonormal 
superpositions of $\nk{n}_t$
\begin{equation}\label{eq:psi}
\psiSkt{i}=\sum_{\vec n|N_\sigma}C^\sigma_{i;\vec n}(t)\nk{n}_t,\quad i=1,...,M
\end{equation}
where symbol '$\vec n|N_\sigma$' indicates that the summation runs over all 
$K_\sigma$ possible number states for an $N_\sigma$-particle system. These 
states are the aforementioned SBSs, on which
the expansion \eqref{eq:PSI} is based.

In summary, the state of the system $\ket{\Psi(t)}$ is expanded 
with respect to the SBSs $\{\psiSkt{i}\}_{i=1}^{M}$ and the SPFs
$\{\phiPkt{i}\}_{i=1}^{m_\sigma}$, which corresponds to a two-layer Hilbert space truncation:
 On the so-called particle layer, the assignment of a set of SPFs to each species truncates the 
single-particle Hilbert space $\mathfrak{h}_n^\sigma$ to a smaller sub-space of dimension $m_\sigma <
G_\sigma$. On the so-called species layer, one takes only a reduced number of $M<\min\{K_A,K_B\}$ 
time-dependent SBSs for each species into account instead of expanding the state of the binary mixture w.r.t.\ 
all conceivable number-state configurations of the two species 
$\nk[A]{n_A}_t\nk[B]{n_B}_t$ with the SPFs as the underlying single-particle 
states. Thereby, $m_\sigma$ and $M$ constitute the main numerical control parameters (see the discussion in 
Sect.\ \ref{sec_limit_cases}).
In the following section, we explain how the optimal time-dependent basis states
are found. For simplicity, we drop the time arguments in the 
notation in the remainder of this work.

\subsection{Equations of motion}
In order to derive equations of motion (EOM) for the expansion 
coefficients $A_{ij}$ and for the time-dependent basis states $\phiPk{i}$, 
$\psiSk{i}$, which ensure that these basis states move in an optimal manner,  
we employ the Langrangian variational principle,  which turns out to 
be technically most convenient \cite{Alon2008,EQ_VarPr}. Thus, 
we aim at finding the stationary point of the action functional (setting $\hbar=1$)
\begin{equation}\label{eq:action}
S =\int_0^t d\tau\,\bra{\Psi(\tau)}\big(\hat 
H-i\partial_\tau\big)\ket{\Psi(\tau)},
\end{equation}
under the constraints of conserving the normalization of the total state 
as well as the orthonormality of the SPFs and SBSs. Enforcing these constraints 
by Lagrange multipliers, however, does not give unique EOM, since expansions
w.r.t.\ time-dependent basis states such as \eqref{eq:PSI} are not unique: 
Time-dependence can be shuffled between the expansion coefficients and the 
time-dependent basis states by means of appropriate (time-dependent) unitary 
transformations of the expansion coefficients and the basis states leaving the 
total state invariant. This ``gauge'' degree of freedom can be fixed by 
setting the so-called
constraint operators $\phiPb{q}i\partial_t\phiPk{k}$ and
$\psiSb{q}i\partial_t\psiSk{k}$ to some hermitian matrices
\cite{MCTDH_BJMW2000}, which we take to be zero in this work without loss of 
generality and without affecting the accuracy of the method. 
Now let us summarize the resulting EOM, while presenting the explicit formulas for the ingredients in 
Appendix \ref{app_trans_mat} - \ref{app_dmat}.

When performing the variation of $A^*_{ij}$, we obtain the 
following EOM
\begin{equation} \label{eq:eom1}
i\partial_t
A_{ij}=\sum_{q,p=1}^{M}\bra{\psiSlabel[A]{i}\psiSlabel[B]{j}}\hat H\ket{
\psiSlabel[A]{q}\psiSlabel[B]{p}}\,A_{qp}.
\end{equation}
As expected, the expansion coefficients obey a linear Schr\"odinger equation
with a, by virtue of the SBSs, time-dependent Hamiltonian matrix. 
Varying correspondingly the coefficients $(C^\sigma_{i;\vec n})^*$ gives
\begin{equation}\label{eq:eom2}
i\partial_t C^\sigma_{i;\vec n}=
\nb{n}(1-\hat P^\sigma_1)\left[\big(
\hat 
H_\sigma+\hat V_\sigma\big)\psiSk{i}+
\sum_{l,j,k,p=1}^{M}
[\eta^{-1}_{1,\sigma}]^i_l\,[\eta_{2,\sigma\bar{\sigma}}]^{lk}_{jp}\,[\hat{W}_{\sigma|\bar{\sigma}}]^k_p\,
\psiSk{j} \right],
\end{equation}
where $\bar{\sigma}=A(B)$ if $\sigma=B(A)$.
The SBSs dynamics is driven by both the intra-species Hamiltonian $\hat H_\sigma+\hat V_\sigma$ and the 
coupling to the other species: the mean-field operator matrix $[\hat{W}_{\sigma|\bar{\sigma}}]^k_p = 
\psiSb[\bar{\sigma}]{k}\hat W_{AB}\psiSk[\bar{\sigma}]{p}$, being an effective one-body 
operator, describes the action of the inter-species interaction on the $\sigma$ species conditioned on the 
transition of the $\bar{\sigma}$-species from the $p$-th to the $k$-th SBS. This 
operator is weighted with the corresponding two-species density 
matrix $[\eta_{2,\sigma\bar{\sigma}}]^{lk}_{jp}$ in the SBS representation
and contracted with the inverse of the reduced density matrix \footnote{\label{foot:rdm_reg}As one can see in 
the spectral representation, the inverse of the 
reduced density matrix makes SPFs of low population rotate fast in oder to become dynamically 
optimized. Note that (near) singular density matrices have to be regularized (see 
\cite{MCTDH_BJMW2000} and the recently proposed alternative to the regularization 
\cite{manthe2015}).} of the $\sigma$ species in the SBS representation, 
$[\eta_{1,\sigma}]^i_l$.
Summarizing, the inter-species
interaction becomes manifest in a two-fold manner: it acts as an effective one-body operator on the $\sigma$ 
species but also leads to a coupling of $i\partial_tC^\sigma_{i;\vec n}$ to the expansion coefficients of 
all the other SBSs.
The operator $1-\hat P^\sigma_1\equiv1-\sum_{r=1}^M\psiSk{r}\!\psiSb{r}$ denotes the 
projector onto the orthogonal complement of the space spanned by the SBSs 
w.r.t.\ ${\rm span}\{\nk{n}_t\}$ naturally preventing the SBSs from moving within the already spanned 
subspace ${\rm span}\{\psiSk{r}\}_{r=1}^M$.

Finally varying the SPFs, we end up with
\begin{align}\label{eq:eom3}
i\partial_t \phiPk{i}=(1-\hat P^\sigma_2)\left[\hat h_\sigma\phiPk{i}
+\sum_{p,s=1}^{m_\sigma}[\rho^{-1}_{1,\sigma}]^i_p\Big(
 \sum_{q,l=1}^{m_\sigma}[\rho_{2,\sigma}]^{pq}_{sl}[\hat v_\sigma]^q_l
+\sum_{q,l=1}^{m_{\bar{\sigma}}}[\rho_{2,\sigma\bar{\sigma}}]^{pq}_{sl}
[\hat w_{\sigma|\bar{\sigma}}]^q_l\Big)\phiPk{s} \right].
\end{align}
Here again, the projector $1-\hat
P^\sigma_2\equiv1-\sum_{i=1}^{m_\sigma}\phiPk{i}\!\phiPb{i}$ enforces that the
SPFs may only rotate out of the subspace spanned by the instantaneous SPFs.
Such a rotation can be induced by the single-particle Hamiltonian $\hat
h_\sigma\equiv\hat h^\sigma_1$
and by the interaction with particles of the same (other) species via
the mean-field operator matrices $[\hat v_\sigma]^q_l$ ($[\hat
w_{\sigma|\bar{\sigma}}]^q_l$). These operators have potential-term
character \footnote{Given that the particles are interacting via
interaction potentials $v_\sigma(\hat x_1^\sigma,\hat x_2^\sigma)$,
$w_{AB}(\hat x_1^A,\hat x_1^B)$, i.e.\ local operators, as it is natural to 
assume.} and are given by
the interaction-potential operators conditioned on the interaction partner
undergoing a transition from the $l$-th to the $q$-th SPF:
$[\hat v_\sigma]^q_l=
\phiPb{q}v_\sigma(\hat x_1^\sigma,\hat x_2^\sigma)\phiPk{l}$ and
$[\hat
w_{\sigma|\bar{\sigma}}]^q_l= \phiPb[\bar{\sigma}]{q}w_{AB}(\hat x_1^A,\hat
x_1^B)\phiPk[\bar{\sigma}]{l}$.
These transitions are weighted with the corresponding
reduced two-body density matrix elements of two
$\sigma$ particles (one $\sigma$ and one $\bar{\sigma}$
particle) in the SPF representation, namely $[\rho_{2,\sigma}]^{pq}_{sl}$ 
($[\rho_{2,\sigma\bar{\sigma}}]^{pq}_{sl}$).
Last, these 
terms are contracted with the
inverse of the reduced one-body density 
matrix \cite{Note4} of a $\sigma$
particle, $[\rho^{-1}_{1,\sigma}]^i_p$.

The coupled EOM \eqref{eq:eom1}, \eqref{eq:eom2}, and \eqref{eq:eom3}, which
have to be solved simultaneously, formally appear alike the EOM of the
ML-MCTDHB method for bosonic mixtures \cite{kronke_non-equilibrium_2013,cao_multi-layer_2013}, which has been 
achieved by the employed unifying notation. As a matter of 
fact, the particle-statistics enter the ML-MCTDHX EOM only via the second
quantization algebra when applying $\hat H_\sigma$, $\hat V_\sigma$ and
$[\hat{W}_{\sigma|\bar{\sigma}}]^k_p$ to a SBS in \eqref{eq:eom2} as well as in
the calculation of the so-called reduced one- and two-body transition matrices
\cite{many_electr_and_red_damats,maziotti_rdm_book07} 
$\psiSb{k}\ad{i}\a{j}\psiSk{l}$ and $\psiSb{k}\ad{i}\ad{j}\a{q}\a{p}\psiSk{l}$
(see Appendix~\ref{app_trans_mat}). 
The latter entities measure the overlap
between the many-body states $\psiSk{l}$, $\psiSk{k}$ after removing one (two)
particles, i.e.\ probe one (two) body properties in the transitions
between  many-body states, and constitute essential building
blocks for constructing the Hamiltonian matrix $\bra{\psiSlabel[A]{i}\psiSlabel[B]{j}}\hat H\ket{
\psiSlabel[A]{q}\psiSlabel[B]{p}}$, the mean-field operator matrices
$[\hat{W}_{\sigma|\bar{\sigma}}]^k_p$ and the reduced density matrices $[\rho_{1,\sigma}]^i_p$,
$[\rho_{2,\sigma}]^{pq}_{sl}$ and $[\rho_{2,\sigma\bar{\sigma}}]^{pq}_{sl}$
(see Appendix \ref{app_hamilt_mat} - \ref{app_dmat}). In 
this way, the reduced transition matrices are reminiscent of the reduced 
density operators in the unified formulation 
\cite{mctdh_for_mixtures_of_2types_of_identical_particles_PRA2007,MCTDHX_recursively} of the 
MCTDH theory for mixtures of indistinguishable 
bosons and / or fermions, which, however, is limited to small particle numbers since the additional 
truncation to $M$ SBSs is not performed.
 
\subsection{Limiting cases and convergence}\label{sec_limit_cases}
Whenever the basis set on one layer is not truncated, the projector $\hat
P^\sigma_1$ or $\hat
P^\sigma_2$ in the corresponding EOM becomes the identity such that the
corresponding basis states do not move (in the chosen constraint operator
gauge). In particular, the ML-MCTDHX EOM reduce to the MCTDH theory for BB, FF and BF mixtures
in its unifying formulation 
\cite{mctdh_for_mixtures_of_2types_of_identical_particles_PRA2007,MCTDHX_recursively}
if $M=\min\{K_A,K_B\}$. When choosing $m_\sigma=G_\sigma$ in addition, 
our wave-function ansatz is of full configuration-interaction type, covering the complete Hilbert space 
$\mathscr{H}$.
In the other extreme case of neglecting all correlations by setting
$M=1$ and $m_\sigma=1$ for bosons ($m_\sigma=N_\sigma$ for fermions),
the ML-MCTDHX EOM reduce to coupled GP (HF) EOM, depending on the statistics of the
components. In between these extreme cases for not too strong inter-species correlations,
one expects that it is sufficient to take only $M\ll \min\{K_A,K_B\}$ variationally optimized SBSs into
account, which leads to higher efficiency in the representation of the
many-body state compared to the direct extension of the MCTDHF and MCTDHB methods for treating mixtures of indistinguishable 
degrees of freedom
\cite{mctdh_for_mixtures_of_2types_of_identical_particles_PRA2007,MCTDHX_recursively}. Thus when using 
ML-MCTDHX, $m_\sigma$ and $M$ need to be chosen according to the
dominant correlations in the system such that the expansion \eqref{eq:PSI} is efficient, 
i.e.\ involves as few coefficients as possible, while capturing all relevant 
correlations. We remark that the EOM 
can be shown to conserve energy and norm \cite{MCTDH_BJMW2000} as well as
single-particle symmetries (see \cite{cao_multi-layer_2013} for details) independently of the chosen 
number of SPFs and SBSs.

In order to judge the convergence of a simulation, one has to systematically 
increase $m_\sigma$ and $M$, i.e.\ enlarge the subspace within which ML-MCTDHX finds the variationally 
optimal solution, and compare the results for observables of interest, e.g.\ the density distribution. 
Furthermore, one can inspect
the spectral decomposition of the reduced density operator of the whole species $\sigma$
\begin{equation}\label{eq_NSF_def}
 \hat\eta_{1,\sigma} = \sum_{i=1}^M\lambda_i^{\sigma}\,|\Psi^{\sigma}_i\rangle\!\langle\Psi^{\sigma}_i|,
\end{equation}
and the reduced one-body density operator of the species $\sigma$
\begin{equation}\label{eq_NO_def}
 \hat\rho_{1,\sigma} = 
\sum_{i=1}^{m_\sigma}n_i^{\sigma}\,|\Phi^{\sigma}_i\rangle\!\langle\Phi^{\sigma}_i|.
\end{equation}
The eigenstates $|\Psi^{\sigma}_i\rangle$ [$\,|\Phi^{\sigma}_i\rangle\,$] are called natural species functions 
(NSFs) [natural orbitals (NOs)] and the eigenvalues $\lambda_i^{\sigma}$ [$\,n_i^{\sigma}\,$], which sum up 
to unity, are denoted as natural populations (NPs). We label the NSFs and NOs in 
decreasing sequence w.r.t.\ their natural populations in the following and note that 
$\lambda_i^{A}=\lambda_i^{B}\equiv\lambda_i$ holds for binary mixtures (see Sect.\ 
\ref{sec_generalization}). 
Convergence can now be judged on the basis of the 
convergence of the individual natural populations. Thereby, a negligibly small lowest natural population 
$\lambda_M^\sigma$ ($n^\sigma_{m_\sigma}$) means that the corresponding NSF (NO) essentially does not 
contribute to the numerically obtained $\ket{\Psi(t)}$. The latter provides in practice a good indicator for 
the considered basis being sufficiently large, even though it is not a strict convergence criterion \cite{Cosme2016}.  
We note that the convergence behavior and in particular the degree of convergence in general depend on
the character of the observable of interest, of course.
In Sect.\ \ref{chap:app}, we will illustrate how to control
convergence of ML-MCTDHX simulations and, moreover, employ the NSFs and NOs as well as their populations as 
an analysis tool for a descriptive unraveling of correlated many-body states.

\subsection{Generalizations}\label{sec_generalization}
Having outlined the ML-MCTDHX wave-function ansatz and EOM for a one-dimensional binary mixture, we briefly 
comment on the possible generalizations, which we can also treat with our current 
implementation. First, more than two species can  straightforwardly be taken into account by employing the 
ansatz
\begin{equation}
 \ket{\Psi}=\sum_{j_1=1}^{M_1}...\sum_{j_S=1}^{M_S}A_{j_1...j_S}\;\psiSk[1]{j_1}\cdots\psiSk[S]{j_S}
\end{equation}
with $S$ denoting the number of species. As a consequence, the calculation of the one- and two-species
density matrix has to be adapted (see Appendix \ref{app_dmat}) and the EOMs for the SBSs and SPFs of the 
$\sigma$ species feature
$S-1$ inter-species mean-field operator matrix terms. While for $S>2$ the numbers of SBSs $M_\sigma$ for the 
different
species do not have to coincide, we remark that taking unequal numbers $M_A$, $M_B$ for a binary mixture does 
not
improve the accuracy of the wave-function ansatz. This is a consequence of the Schmidt decomposition 
for bipartite systems \cite{nielsen_chuang_book,Horodecki}, which states that the maximal strength of 
interspecies correlations, 
which can be resolved, only depends on $\min\{M_A,M_B\}$ such that we may set 
$M_A=M_B=M$ without loss of generality. This feature is also reflected in the aforementioned coinciding 
natural
populations of the NSFs, $\lambda_i^{A}=\lambda_i^{B}\equiv\lambda_i$.

Second, the concept of multi-layering can be applied to simulate also three-dimensional systems. Here, the 
idea is to expand the time-dependent three-dimensional SPFs $\phiPk{i}$ w.r.t.\ Hartree products of 
time-dependent one-dimensional basis states for the different spatial directions. By adjusting the numbers of 
these one-dimensional basis states, correlations between the spatial directions can be taken into account. If 
the latter are not too strong, as in the physically 
interesting cross-over regime between one and three dimensions, this approach allows for efficient simulations
on extremely large spatial grids \cite{cao_multi-layer_2013,Bolsinger2017,bolsinger_scattering_hump}.
In a completely similar manner, we can account for internal degrees of freedom in ML-MCTDHX, allowing
for simulating in particular interacting ultracold fermions in different pseudo-spin states including also 
spin-changing collisions \cite{krauser_giant_2014}. For this purpose, the SPFs $\phiPk{i}$ are expanded 
w.r.t.\ Hartree products of time-dependent spatial wave-functions and time-independent spin basis states.


\section{Applications: Collisional dynamics of binary mixtures}
\label{chap:app}

In this section, we apply the ML-MCTDHX method for simulating the correlated collision dynamics of various 
binary mixtures in order to demonstrate the versatility and certain features of the approach. After discussing 
the general setup in Sect. \ref{sec:Hamiltonian}, we focus on the convergence behavior 
by considering a simple scenario involving a few-atom BF mixture (Sect. \ref{sec:bench1}). Thereafter,
we proceed to more complex scenarios involving larger atom numbers and illustrate how manifestations of 
correlations can be unraveled by means of tailored analysis tools. Here, we consider the cases of a 
particle-number imbalanced BF (Sect. \ref{sec:bench2}) and a balanced FF mixture (Sect. \ref{sec:bench3}). We 
note that exploring the observed physical effects in depth clearly goes beyond the scope of this work. 

\subsection{Setup}
\label{sec:Hamiltonian}
We consider a binary mixture consisting of $A$ and $B$ species of the same mass \footnote{For simplicity, we 
consider 
the constituents to possess the same mass also 
for the case of a BF mixture.},     
being confined in harmonic-oscillator potentials of same frequencies. Working in harmonic oscillator units, the single-particle 
Hamiltonian takes the form 
$h_i^{\sigma}=-\frac{1}{2}\partial^2_{x_{i}^\sigma}+\frac{1}{2}(x_i^{\sigma}-x_0^{\sigma})^2$ where 
$x_{0}^{\sigma}$ (${\sigma} \in \lbrace A,B \rbrace$ is the species index) denotes the spatial offset added 
initially to the harmonic confinement for each of the species. The two-body intra-species and 
inter-species interaction potential is given by 
$v_{\sigma}(x_i^{\sigma},x_j^{\sigma})=g_{\sigma}\delta(x_i^{\sigma}-x_j^{\sigma})$ and 
$w_{AB}(x_i^{A},x_j^{B})=g_{AB}\delta(x_i^{A}-x_j^{B})$ 
respectively.  

Below, we explore the dynamics of both a BF and a FF mixture with each of the individual species 
being considered as spinless \footnote{We remark here that even one of the species would possesses spin degrees of freedom they are 
considered to be frozen as is the case of a spin polarized ensemble.}.  
Each system is initialized in the ground state of the 
corresponding species with an initial finite spatial offset $x_{0}^{A}=-x_{0}^{B}$. 
To induce the dynamics, we quench this spatial offset to zero, i.e. $x_{0}^{A}=-x_{0}^{B}=0$, 
letting the initially separated atomic clouds collide. 

\subsection{Few-atom Bose-Fermi mixture: Convergence study}
\label{sec:bench1} 

In the present section, we consider the dynamics of a few-body BF mixture and show in detail the  
convergence behavior of the ML-MCTDHX method. In particular, the mixture consists of $N_B=2$ bosons and 
$N_F=2$ fermions with weak bosonic intra-species repulsive interaction $g_{BB}=0.05$ and strong repulsive 
inter-species interaction $g_{BF}=1.0$, while the initial spatial offset is $x_{0}^{F}=-x_{0}^{B}=2$. 

\begin{figure}[ht]
\includegraphics[width=0.47\textwidth]{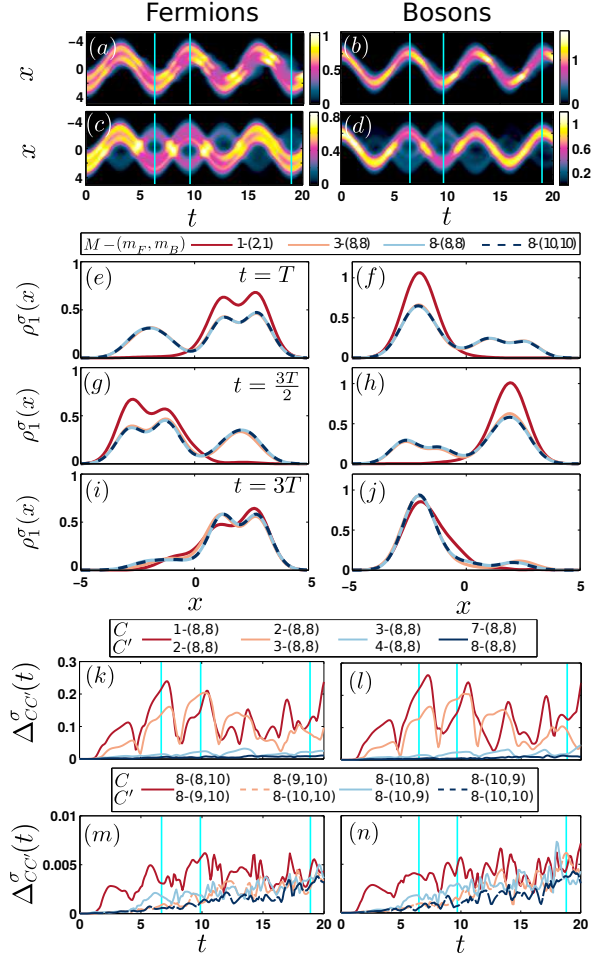}
\caption[]{One-body density evolution $\rho^{\sigma}_1(x;t)$ for the fermions (left column) and the 
bosons (right column) after the quench for a system of $N_F=2$ fermions and $N_B=2$ bosons and
with intra-species and inter-species interaction strength $g_{BB}=0.05$ and $g_{BF}=1.0$, respectively.
Shown are the one-body densities evolution ($a,b$) in 
mean-field approximation [$1-(2,1)$] and ($c,d$) for a beyond mean-field 
calculation [$10-(10,10)$].   
($e-j$) Comparison of the one-body density profiles for different orbital configurations (see legend) at 
different time instants, namely from top to bottom $t_1=6.3\approx T$, $t_2=9.4\approx3T/2$ and            
$t_3=18.9\approx3T$. ($k-n$) Spatially integrated error $\Delta_{CC'}^{\sigma}(t)$ between different orbital 
configuration with increasing number of NSFs ($k,l$)  and increasing number of 
SPFs ($m,n$). The light blue vertical lines indicate the time instants shown in $(e-j)$. } 
\label{fig:1}
\end{figure}
To study the dynamics we employ the evolution of the reduced one-body density for the ${\sigma}$-th species 
\be
    \rho^{\sigma}_1(x;t)= \braket{ \Psi(t) | \hat{\Psi}_\sigma^\dagger(x) \hat{\Psi}_\sigma(x) | \Psi (t) 
},\label{eq:1}
\ee 
where $\hat{\Psi}_\sigma^\dagger(x)$ ($\hat{\Psi}_\sigma(x)$) refers to the field operator 
that creates (annihilates) a $\sigma$ species particle at position $x$. 
Figs. \ref{fig:1} ($a$)-($d$) present the dynamics from a single particle perspective via the 
one-body density for both the fermionic and the bosonic species. 
As shown $\rho^{\sigma}_1(x;t)$ exhibits oscillatory patterns possessing the frequency $\omega=1$ of the 
harmonic trap such that at regular time intervals, $\Delta t=\frac{(2n+1)T}{4}$ ($T=2 \pi$ being a 
corresponding period) the two species collide. In the mean-field case, see Figs. \ref{fig:1} ($a$), ($b$), the 
only effect observed is the accumulation of the single particle density before a collision event for both the 
fermionic and bosonic species. However in the correlated case, see Figs. \ref{fig:1} ($c$), ($d$), the 
out-of-equilibrium dynamics differs significantly. As it can be seen, the density distribution $\rho^F_1(x;t)$ 
(as well as $\rho^B_1(x;t)$) oscillates as a whole up to the first collision event (see $t=\frac{1}{2}T$) and 
then splits into counter-propagating fragments. In addition, during each collision event the density shows a 
characteristic peak, which is more pronounced than in the mean-field case. The observed splitting process can 
be seen as a partial reflection of the particles during a collision. Furthermore note that the reflected 
part of the one-body density lies within the spatial region for which the density of the other species is dominant 
and inherits its density structure. 

Since beyond mean-field effects apparently play a role already on the level of the reduced one-body density, 
we first study the convergence behavior of $\rho^{\sigma}_1(x;t)$ upon increasing the basis size in our 
variational wave-function ansatz [see Eqs. (\ref{eq:PSI}), (\ref{eq:psi})]. Figs. \ref{fig:1} ($e$)-($j$) show the 
profiles of the one-body density for both the fermions and the bosons using different $M-(m_F,m_B)$ 
approximations at various time instants during the evolution. For increasing number of SBSs $M$ (keeping the 
number of SPFs $m_B$, $m_F$ fixed) the corresponding profiles are almost indistinguishable (see solid red and 
light blue lines respectively). For instance, the corresponding one-body densities obtained e.g. within the 
approximations 3-(8,8) and 8-(8,8) respectively, deviate at most by $10\%$. 
However, the difference between the 7-(8,8) and 8-(8,8) approximations becomes negligible, 
showing a maximal deviation of the order of $1.2\%$ (results not shown here).  
The same behavior can be observed for increasing number of SPFs, namely by increasing $m_B$ and/or $m_F$ for a 
fixed number of NSFs $M$ (see light blue and black dashed lines). Here the maximum deviation between the 
approximations 8-(8,8) and 8-(10,10) is of the order of $2\%$.  However, the mean-field approximation 1-(2,1) 
differs significantly from the correlated approach 8-(10,10) possessing completely different density profiles 
(see solid red and dashed black lines). 

In order to quantitatively judge the convergence behavior of $\rho^{\sigma}_1(x;t)$ over the whole evolution 
time, we inspect the spatially integrated density difference 
\be
    \Delta_{C C'}^{\sigma}(t)=\frac{1}{2N_{\sigma}}\int dx 
\abs{\rho^{\sigma}_{1,C}(x;t)-\rho^{\sigma}_{1,C'}(x;t)}\in [0,1],\label{eq:2} 
\ee
where $N_{\sigma}$ refers to the total number of particles for the species $\sigma$, 
$\rho^{\sigma}_{1,C}(x;t)$ ($\rho^{\sigma}_{1,C'}(x;t)$) denotes the one-body density of 
species $\sigma$ calculated within the configuration $C=M-(m_F,m_B)$ [$C'=M'-(m^{'F},m^{'B})$]. 
To judge about convergence on the single-particle level we calculate $\Delta_{C 
C'}^{\sigma}(t)$ by successively increasing $M$ and fixed $m_F$, $m_B$ or with successively increasing the 
$m_B$ or $m_F$ keeping $M$ fixed. 

Figs. \ref{fig:1} ($k$), ($l$) show in a transparent way an adequate convergence of our results both for 
fermions [see Fig. \ref{fig:1} ($k$)] and bosons [see Fig. \ref{fig:1} ($l$)] for increasing number of SBSs 
and in particular for $M>7$. Indeed, $\Delta_{CC'}^{\sigma}(t)$ testifies large deviations for small number 
of SBSs (e.g. incrementing from $M=1$ to $M=2$), while for increasing $M$ (here $M=7$ to $M=8$) 
$max[\Delta_{CC'}^{\sigma}(t)]\approx 1\%$.  The latter indicates convergence of the reduced one-body 
densities.  The same observations hold for increasing number of SPFs in the fermionic and/or the bosonic 
component of the binary mixture. As shown in Figs. \ref{fig:1} ($m$), ($n$) the one-body densities for both 
the fermions and the bosons respectively  are more sensitive to the addition of more orbitals $m_F$ in the 
fermionic component. This is an expected result as spinless fermions, in contrast to bosons, according to 
the Pauli principle cannot tend to condense to a particular state.  An adequate convergence for increasing 
number of fermionic orbitals between the configurations $C=8-(8,10)$, $C'=8-(9,10)$ (see solid red line) and 
$C=8-(9,10)$, $C'=8-(10,10)$ (see dashed orange line) is observed.  In particular, 
$\max[\Delta_{CC'}^{\sigma}(t)] \approx 0.5\%$ (see solid light blue and blue dashed lines) for $m_F,m_B>9$. 
Summarizing, we observe that the accuracy of the calculation is more sensitive upon increasing the number of 
SBSs than incrementing the number of SPFs. Also, to achieve convergence we need to add more SPFs for the 
fermions than for the bosons which can be explained by the fact that bosons are allowed to condense in the 
same state but fermions not.  
\begin{figure*}[ht]
\includegraphics[width=0.95\textwidth]{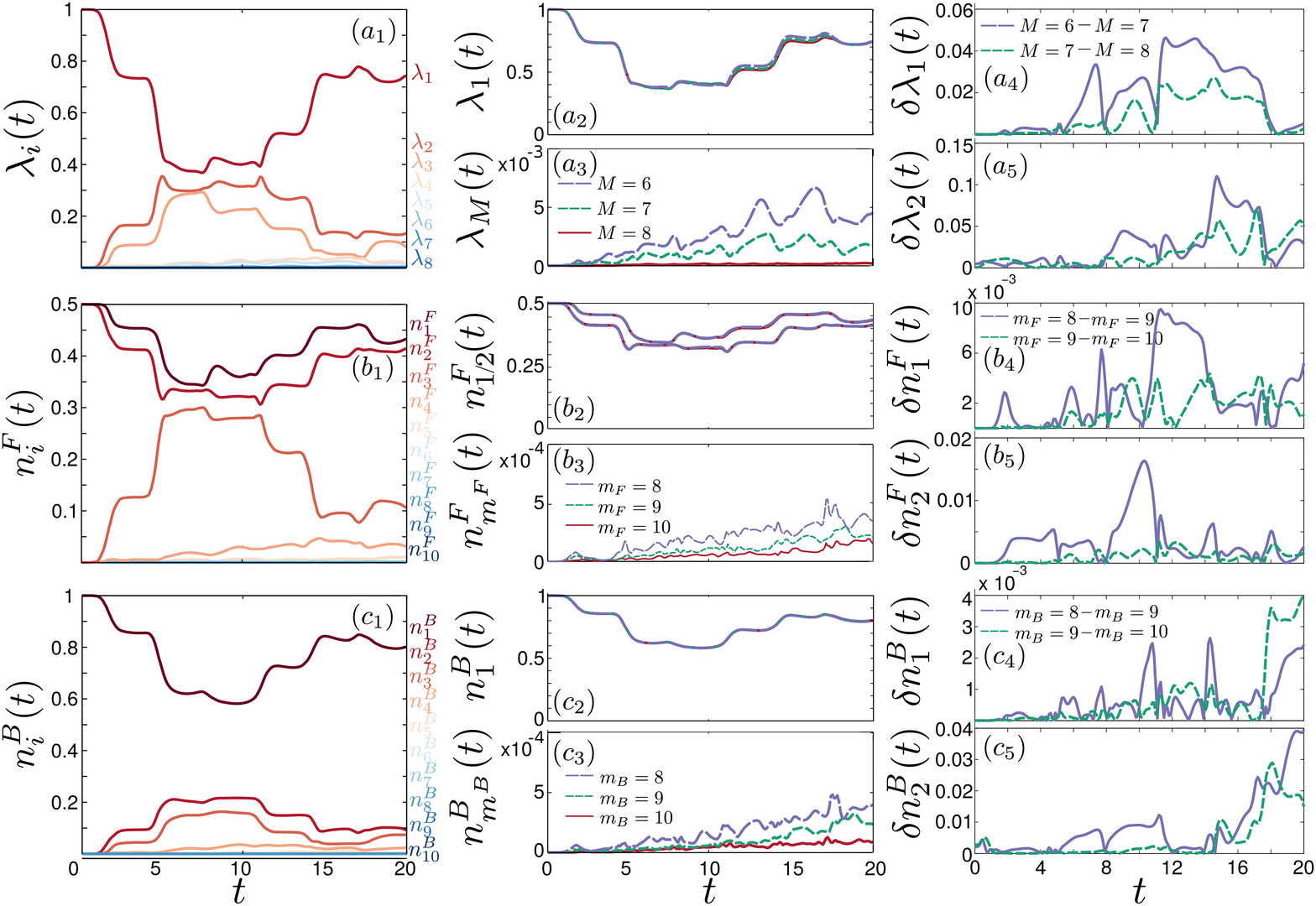}
\caption[]{Evolution of the occupations of the NSFs ($a_1$), the fermionic NOs ($b_1$), and the bosonic NOs 
($c_1$) for the orbital configuration $8-(10,10)$.
Comparison of the populations of the  most occupied and least occupied NSF (NO) for 
varying $M$ with fixed $m_F=m_B=10$ ($a_2,a_3$), for varying $m_F$ with fixed $M=8$ and $m_B=10$ ($b_2,b_3$), 
and for varying $m_B$ with fixed $M=8$ and $m_F=10$ ($c_2,c_3$).
Corresponding relative differences for the natural population of the first and second NSFs ($a_4,a_5$) 
and NOs ($b_4,b_5$ and $c_4,c_5$) between the different orbital configurations (see legend).}
\label{fig:2}
\end{figure*}

Having investigated the convergence behavior of local one-body observables, we now turn to more involved 
quantities, namely the populations of NOs and NSFs, which reflect certain aspects of the structure of the 
many-body state (see Sec. II D). The occupations of the NSFs and NOs are important for both the interpretation 
of the dynamics as well as to check the convergence of the method. Fig. \ref{fig:2} ($a_1$) presents the 
evolution of the populations of the NSFs for the system examined in Fig. \ref{fig:1}.  
As it can be seen, $\lambda_1(t=0)=1$, implying the absence of inter-species correlations initially, 
while as time evolves $\lambda_1(t)$ deviates significantly from 
unity and $\lambda_i(t)$, $i\neq 1$, acquire non-negligible populations.    
Indeed all $\lambda_i(t)$, $i=1,2,...$, change significantly during a collision event but stay approximately constant  
in between.   Figs. \ref{fig:2} ($a_2$), ($a_3$) illustrate in a transparent way, the convergence regarding 
the first and last NSF population respectively for varying number of used SBSs, namely $M=6,7,8$.  
Concerning the occupation of the least contributing NSF for the case of $M=8$ we observe that it is always very small, i.e. less than $10^{-3}$. 
The latter indicates that the used number of SBSs is sufficient for the convergence of the observables of interest (see
Sect.\ \ref{sec_limit_cases}).
The occupations of the first NSF during the evolution for different orbital configurations are 
almost indistinguishable. To quantitatively demonstrate convergence we define the relative 
difference of the population of the $i$-th NSF  
\be
\delta\lambda_i^{CC'}(t)=\frac{\abs{\lambda_i^{C}(t)-\lambda_i^{C'}(t)}}{\lambda_i^{C'}(t)},\label{eq:3} 
\ee
where $\lambda_i^{C}(t)$ [$\lambda_i^{C'}(t)$] denotes the population of the $i$-th NSF calculated within the 
$C=M-(m_F,m_B)$ [$C'=M'-(m^{'F},m^{'B})$] orbital configuration.  Figs. \ref{fig:2} ($a_4$), ($a_5$) present  
the relative difference for the first $\delta\lambda_1^{CC'}(t)$ and the second $\delta\lambda_2^{CC'}(t)$ 
population of the NSFs. A decreasing relative difference for increasing $M$ is clearly visible
and the relative deviations are very small anyway, i.e.\ less than $4\%$
when comparing the $C=7-(10,10)$ calculation with $C'=8-(10,10)$, for example, which
further indicates convergence.

Moreover, Fig. \ref{fig:2} ($b_1$) presents the natural populations (NPs) of the fermionic natural orbitals during the 
dynamics. As $N_F=2$, the fermionic ensemble initially occupies a HF-like state characterized by the 
occupation of the first two orbitals, i.e. $n_1^F(0)=n_2^F(0)\approx0.5$. 
However, following the first collision, other fermionic 
orbitals acquire significant 
populations. 
To judge about the obtained convergence Figs. \ref{fig:2} ($b_2$), ($b_3$) show the evolution of the first two and of 
the last fermionic NPs respectively using different number 
of fermionic orbitals $m_F$. Concerning the dependence of $n_1^F(t)$, $n_2^F(t)$ on the numerical 
configuration, we observe for the presented orbital numbers that
they are almost indistinguishable during the evolution, see Fig. \ref{fig:2} ($b_2$). The latter is 
also confirmed by their corresponding relative differences $\delta n_1^{F,CC'}(t)$ and $\delta 
n_2^{F,CC'}(t)$ [defined in the same manner as the relative difference of 
the population of the NSFs, see Eq.(\ref{eq:3})], shown in Figs. \ref{fig:2} ($b_4$), ($b_5$) respectively.  A 
decaying relative difference for increasing number of fermionic orbitals [see $\delta n_1^{F,CC'}(t)$ for 
$C=8-(8,10)$ and $C'=8-(9,10)$ and $\delta n_1^{F,CC'}(t)$ between  
$C=8-(9,10)$ and $C'=8-(10,10)$] is observed with a corresponding deviation smaller than $0.6\%$, thereby 
indicating convergence. Finally, the smallest natural population, see Fig. \ref{fig:2} ($b_3$), is  
less than $10^{-3}$, indicating that there is no necessity of adding more orbitals to the 
fermionic component. Next we turn our attention to the evolution of the bosonic NPs, shown in Fig. \ref{fig:2} 
($c_1$). The same overall behavior as for the fermionic NPs is observed. 
To illustrate convergence we explicitly show in Figs. \ref{fig:2} ($c_2$), ($c_3$) the evolution of the first and the last NP for 
varying number of bosonic NOs $m_B$. The time-evolution of the first NP, see Fig. \ref{fig:2} ($c_2$), remains 
almost invariant upon increasing the orbital numbers, while the last one, see Fig. \ref{fig:2} ($c_3$), is 
always below $10^{-3}$. Finally, the corresponding relative differences of the first [$\delta n_1^{B,CC'}(t)$] 
and the second [$\delta n_2^{B,CC'}(t)$] bosonic NPs using different number of $m_B$, are presented in Figs. \ref{fig:2} ($c_4$), ($c_5$). 
As expected, both $\delta n_1^{B,CC'}(t)$ and $\delta n_2^{B,CC'}(t)$ decrease 
for increasing number of bosonic NOs $m_B$ (see purple and green lines).  

Summarizing, we have studied the convergence behavior of  the reduced one-body density and the NOs and NSF 
populations, which are sensitive to both intra- and inter-species correlations. 
For the considered configurations, we have in particular seen that the accuracy is much more affected by the 
truncation on the species layer than on the particle layer. We have made similar observations in the other 
collision scenarios, too.

\subsection{Particle-number imbalanced Bose-Fermi mixture} 
\label{sec:bench2}

In the following, we investigate a complex collision scenario where again, for strong enough 
inter-species interactions, correlations 
qualitatively alter the transmission properties of the colliding species.
Here, our main aim is to show that the NSFs, which can be easily extracted from the ML-MCTDHX wave-function 
ansatz, allow for linking the structure of the correlated many-body wave function to beyond mean-field effects 
in the reduced one-body density. 
Specifically, we consider a BF mixture consisting of $N_B=10$ bosons and 
$N_F=3$ fermions, while the initial positions 
of the species-dependent harmonic trap are $x_{0}^{F}=-x_{0}^{B}=2$.  
The bosonic intra-species interactions is kept always weak ($g_{BB}=0.05$) and the cases of weak and stronger  
inter-species interactions, 
namely $g_{BF}=0.5$ and $g_{BF}=2.0$ respectively, are examined. 
\begin{figure}[ht]
\includegraphics[width=0.50\textwidth]{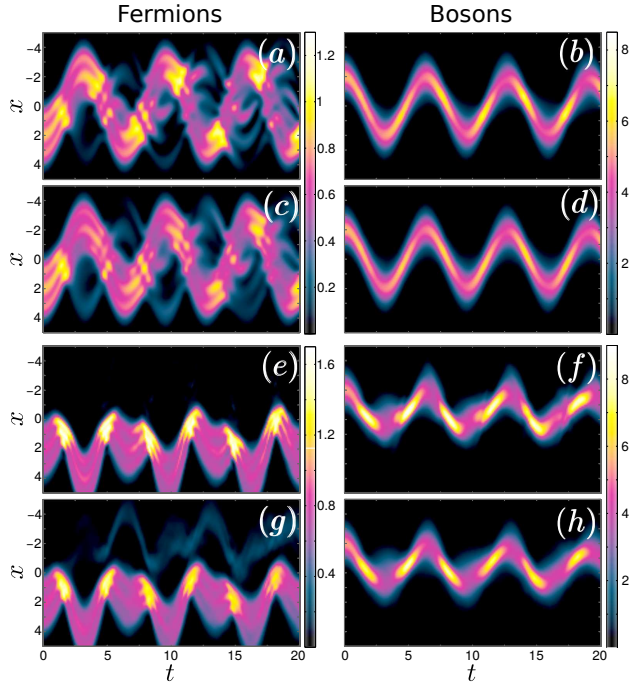}
\caption[]{One-body density evolution for the fermionic species (left column) and the bosonic species (right 
column) after the quench for $N_F=3$ fermions and $N_B=10$ bosons with 
intra-species interaction $g_{BB}=0.05$ and inter-species interactions $g_{BF}=0.5$ ($a-d$) and 
$g_{BF}=2.0$ ($e-h$). While densities shown in panels ($a,b,e,f$) are obtained within a mean-field 
configuration [$1-(3,1)$] panels ($c,d,g,h$) present the beyond mean-field results [$6-(8,6)$].} \label{fig:3}
\end{figure}

\begin{figure}[ht]
\includegraphics[width=0.50\textwidth]{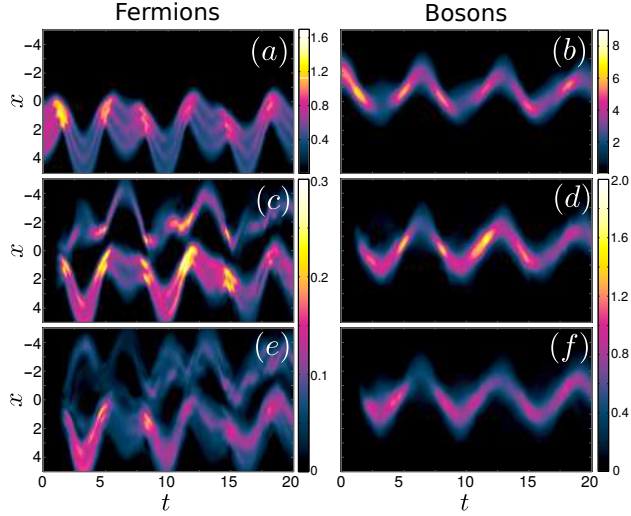}
\caption[]{Evolution of the leading contribution $\lambda_1(t) 
\rho^{\sigma}_{1,1}(x;t)$  ($a,b$) , the next-to-leading contribution $\lambda_2(t) 
\rho^{\sigma}_{1,2}(x;t)$ ($c,d$)  and the re-summation of the remaining contributions 
$\sum\limits_{k>2} \lambda_k(t) \rho^{\sigma}_{1,k}(x;t)$ ($e,f$) of the one-body density for the fermions 
(left column) and the bosons (right column) for $g_{BF}=2.0$. The remaining system parameters are the same as 
in Fig. \ref{fig:3}.} \label{fig:4}
\end{figure}

For weak inter-species interactions, an inspection of the dynamics of the one 
body density reveals qualitatively similar characteristics between the mean-field,  
[see Figs. \ref{fig:3} ($a$), ($b$)] and the correlated [see Figs. \ref{fig:3} ($c$), ($d$)] description. 
In both cases the two initially separated clouds collide after the quench and perform oscillations with the 
frequency of the trap. The collisional events show in the one-body density evolution density 
peaks. These density peaks appear to be more pronounced in the mean-field case.    
Additionally, as it can be seen after the first collision a small fragment of the fermionic cloud is swept 
away from the majority cloud as a consequence of the collision with the bosonic ensemble.   
For times following the first collision, the bosonic density lies in between the above-mentioned majority 
fermionic cloud and the small reflected fermionic fragment. 
Also, the sweeping of the minority fermionic fragment is intensified in the course of the dynamics especially 
following each collision event. The latter is stronger in the correlated dynamics as compared to the mean-field behaviour.  

Figs. \ref{fig:3} ($e$)-($h$) present the dynamics for stronger interactions, namely $g_{BF}=2.0$. 
In this case the evolution of the system within the mean-field approach is qualitatively different from the 
full many-body calculation. 
Indeed, in the mean-field case, the two species are mainly reflected, see Figs. \ref{fig:3} ($e$), ($f$). 
Only a tiny portion of the fermionic cloud transmits through 
the bosonic ensemble after 
each collision [not visible in Fig. \ref{fig:3} ($e$)]. Taking correlations into accout, however, one part of the 
fermionic density goes through the bosonic species after the  first and even more after the second  
collision, see Figs. \ref{fig:3}($g$), ($h$), and acquires more population during the dynamics.

In order to trace back the transmitted fermionic density to beyond HF behavior, one can evaluate 
the NOs densities and their populations. Being strongly averaged quantities however, the NOs
do not allow for distinguishing intra- and inter-species correlation effects. As we shall see here, the latter 
can be accomplished for a binary mixture by the Schmidt decomposition \cite{Horodecki,Sven}
\be
\ket{\Psi(t)}=\sum_{k=1}^M \sqrt{\lambda_k(t)} \ket{\Psi_k^A(t)} \ket{\Psi_k^B(t)}, \label{eq:4}
\ee 
with $\{\ket{\Psi_k^\sigma(t)},\;k=1,...,M\}$ being orthonormal. This decomposition can be easily obtained by 
diagonalizing the single-species reduced density matrices $[\eta_{1,\sigma}]^i_l$, 
which are typically compact objects in ML-MCTDHX calculations. 
Then, the one-body reduced density $\rho^{\sigma}_{1}(x;t)$ of the $\sigma$ species can be decomposed as  
\be
    \rho^{\sigma}_{1}(x;t)= \sum_{k=1}^M \lambda_k~ \rho^{\sigma}_{1,k}(x;t), \label{eq:5}
\ee
where $\rho^{\sigma}_{1,k}(x;t)=\braket{\Psi^{\sigma}_{k}(t) | \hat{\Psi}_\sigma^\dagger(x) 
\hat{\Psi}_\sigma(x) | \Psi^{\sigma}_{k}(t) }$ denotes the one-body density matrix for the $k$-th NSF.  
Indeed, the decomposition of the one-body density into $\lambda_k(t) \rho^{\sigma}_{1,k}(x;t)$ modes [see Eq. (\ref{eq:5})] 
sheds light on the correlated dynamics and the corresponding differences to the mean-field 
case. The leading contribution $\lambda_1(t) \rho^F_{1,1}(x;t)$, see Figs. \ref{fig:4} ($a$), ($b$), is 
similar to the mean-field one-body density $\rho^F_{1,{\rm MF}}(x;t)$ as the fermionic density is shown to be 
completely reflected after each of the collisions. On the contrary, $\lambda_2(t) \rho^F_{1,2}(x;t)$ clearly 
shows that a small fragment of the fermions is transmitted through the bosonic cloud and the remaining 
fermionic cloud is reflected, see Figs. \ref{fig:4} ($c$), ($d$). Furthermore, in $\sum_{k>2} \lambda_k(t)  
\rho^F_{1,k}(x;t)$ we also observe that a fermionic fragment  
passes through the bosonic cloud, see Figs. \ref{fig:4} ($e$), ($f$). 
For the bosonic component, we do not find striking structural differences between 
the NSF densities but only the tendency that less populated NSFs are more smeared 
out in space, cf.\ $\sum_{k>2} \lambda_k(t) \rho^B_{1,k}(x;t)$ with
$\lambda_1(t) \rho^B_{1,1}(x;t)$ and $\lambda_2(t) \rho^B_{1,2}(x;t)$.

Concluding, we have linked  the transmitted fermionic density to inter-species correlations by diagonalizing 
the density operator of the fermionic species. Because of inter-species interactions, the state of the $N_F$ 
fermions becomes mixed. Its dominant NSF density lacks a transmitted fraction and resembles the behavior of the 
corresponding mean-field calculation, whereas the sub-dominant NSFs feature a transmitted fraction.

\subsection{Fermi-Fermi mixture} \label{sec:bench3}

Finally, we study the collision dynamics of a FF mixture for a two-fold purpose. We can relate - this time - a 
reflected fraction of the density to inter-species correlations by applying the already introduced NSF 
analysis. Furthermore, a NO analysis reveals that the occupied single particle states can be classified into 
certain categories, manifesting the emerging NO structure of the beyond HF dynamics.  

To illustrate the emergence of inter- and intra-species correlations in fermionic ensembles we consider a 
binary mixture consisting of $N_{A}=N_{B}=6$ fermions and $x_{0}^{A}=-x_{0}^{B}=4.5$
in order to avoid spatial overlap between the species initially. We emphasize 
that the considered inter-species interactions are relatively weak, namely $g_{AB}=0.4$,
but still give rise to intriguing correlation effects. 

\begin{figure}[ht] 
\includegraphics[width=0.5\textwidth]{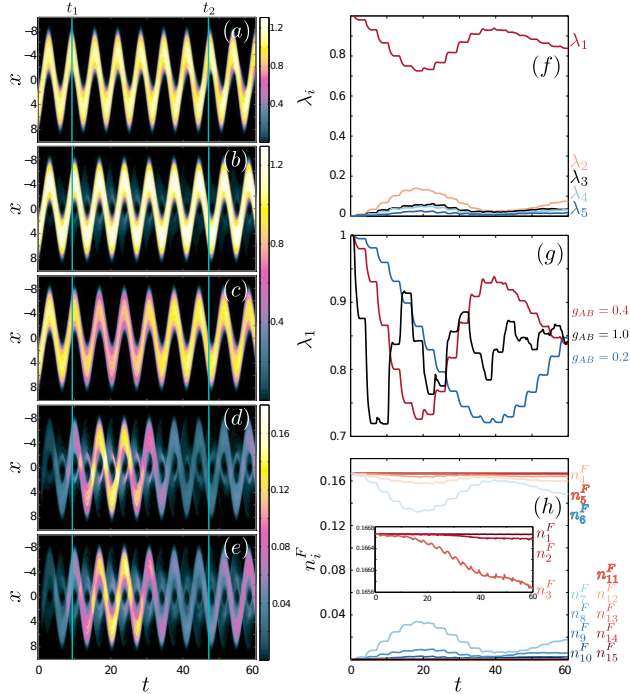} 
\caption[]{Evolution of the one-body 
density for the fermionic species $A$ for $N_{A}=N_{B}=6$ fermions with inter-species interaction 
$g_{AB}=0.4$ within the mean-field approximation ($a$) and resulting from the correlated approach ($b$) with 
the orbital configuration $5-(15,15)$.
($c$) Leading contribution $\lambda_1(t) \rho^{A}_{1,1}(x;t)$, ($d$) next-to-leading contribution 
$\lambda_2(t) \rho^{A}_{1,2}(x;t)$ and ($e$) re-summation of higher order contributions $\sum\limits_{k>2} 
\lambda_k(t) \rho^{A}_{1,k}(x;t)$ to the one-body density of ($b$).
($f$) Evolution of the corresponding populations $\lambda_j$ of the NSFs.
($g$) Comparison of the population for the first NSF during the dynamics for different values of the 
inter-species interaction $g_{AB}=0.2$ (blue), $g_{AB}=0.4$ (red) and $g_{AB}=1.0$ (black). 
($h$) Evolution of the fermionic NPs for $g_{AB}=0.4$. Inset: For better visibility we depict solely the 
populations of the first three fermionic orbitals.}\label{fig:5} 
\end{figure}

\begin{figure}[ht] 
\includegraphics[width=0.5\textwidth]{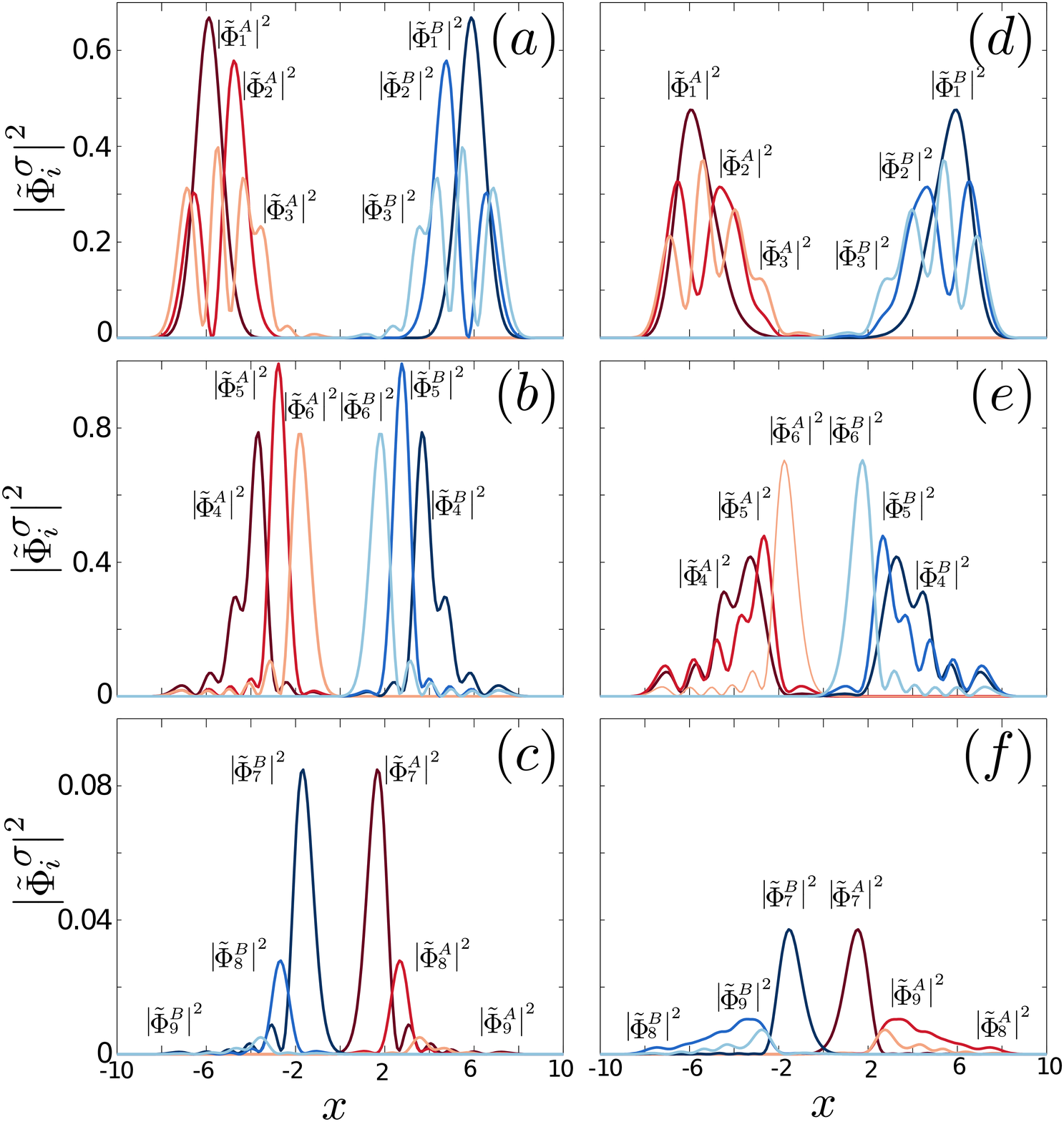} 
\caption[]{Modulus squared of the NOs being normalized to their respective occupation number, i.e.\ 
$\tilde\Phi_i^{\sigma}(x,t)\equiv \sqrt{N_{\sigma} n_i^{\sigma}}\, \Phi_i^{\sigma}(x,t)$ for the orbital 
configuration $5-(15,15)$ and $g_{AB}=0.4$. 
($a,d$) Core orbitals, ($b,e$) valence orbitals, and ($c,f$) excited orbitals at times $t_1=9.8$ (left 
column) and $t_2=46.9$ (right column). The remaining parameters are as stated in Fig. \ref{fig:5}.} 
\label{fig:6}
\end{figure}

The effect of the inter-species interaction is manifested after a collision by a small portion of the 
fermionic density which oscillates in the opposite direction with respect to the majority of the fermionic 
particles, see Fig. \ref{fig:5} ($b$). This is in contrast to the mean-field HF case, see Fig. \ref{fig:5} 
($a$), where no such effect is observed and only coherent oscillations of the fermions are present. To shed 
light onto the collisional fermionic dynamics we employ the corresponding Schmidt decomposition. The density 
evolution of the dominant NSF [see Fig. \ref{fig:5} ($c$)], shows that all the fermions oscillate coherently 
in the same direction, in agreement to the HF case.  The second dominant NSF [see the density evolution in 
Fig. \ref{fig:5} ($d$)] consists of the majority of the fermionic density oscillating in the one direction 
and a small fragment in the opposite direction.  Higher order NSF contributions correspond to states with a 
somewhat more extended density oscillating in the opposite direction [see Fig. \ref{fig:5} ($e$)]. 

The populations of the above-mentioned modes, presented in Fig. \ref{fig:5} ($f$), change over time in particular after each 
collision event (taking place at $t_n=\frac{(2 n + 1) \pi}{2}$, $n \in \mathbb{N}$) and exhibit an oscillatory
stepwise pattern. To compare the depletion rate for the first NSF, i.e.\ the mode where all the particles 
oscillate around the center of the trap in the same direction, we show its occupation for different 
inter-species interactions $g_{AB}$, see Fig. \ref{fig:5} ($g$). It is manifest that stronger
interactions lead to a faster built-up of inter-species correlations. 
Since reflected density fractions are exclusive signatures of higher order NSFs [compare Figs. \ref{fig:5} ($b$), ($d$), ($f$)], 
measuring the density profile can be used to probe inter-species correlations.  
Moreover, the interaction dependence observed in Fig. \ref{fig:5} ($g$) can 
be explained by the fact that the reflection rate for the scattering between each pair of particles gets 
higher for increasing interaction and therefore the probability of all the particles to be transmitted 
decreases. Another interesting observation is that the depletion ($1-\lambda_i$) reaches a maximum value and 
subsequently reduces, e.g. see Figs. \ref{fig:5} ($d$), ($g$). This is due to the collision process between 
the two species which also allows for population transfer from the second NSF to the first.

Similarly to the bosonic case, where the populations of the NOs quantify the degree of fragmentation \cite{Penrose,Mueller},
NOs and their populations provide also valuable insights into the structure of the many-body state of
interacting fermions.  Due to the 
fermionic statistics an amount of the initially occupied orbitals are almost unperturbed (core orbitals) 
during the dynamics, while the remaining orbitals (valence orbitals) exhibit depletion towards higher lying 
excited orbitals not populated in the initial state. In the present case, see Fig. \ref{fig:5} ($h$), 
we identify (from the corresponding populations) the first three $\Phi_1$ to $\Phi_3$ 
orbitals as the core orbitals, the orbitals $\Phi_4$ to $\Phi_6$ as the valence orbitals, and finally the 
orbitals $\Phi_7$ to $\Phi_9$ as the significantly occupied excited orbitals. The core orbitals are 
characterized by low depletion ($1/N_{\sigma} - n_i^{\sigma}$), as shown in the inset of Fig. \ref{fig:5} 
($h$), while the depletion of the valence orbitals is related to the occupation of the excited 
orbitals. Figs. \ref{fig:6} ($a$)-($c$) present $\abs{\Phi_i(x,t_1)}^2$ with $i \in \{1,\dots,9\}$ at 
$t_1=9.8 \approx \frac{3 \pi}{2}$ where the two species have maximal separation. The core orbitals possess a 
similar appearance as the three lowest-energy eigenstates of the harmonic oscillator [visible in the nodal 
structure in Fig. \ref{fig:6} ($a$)]. On the contrary, the three valence orbitals [see Fig. \ref{fig:6} 
($b$)] of each species are closer to each other 
than the core orbitals. Finally, the excited orbitals [see Fig. \ref{fig:6} ($c$)] of one species are located in 
the spatial region where the density of the other species is dominant. Hence, the valence orbitals of 
a given species overlap with the excited orbitals of the other species. The latter reveals that 
the valence orbitals become more affected than the core orbitals by the inter-species interaction. It is 
important to remark that the same kind of orbital structure persists during the evolution, see for instance 
Figs. \ref{fig:6} ($d$)-($f$) where $\abs{\Phi_i(x,t_1)}^2$ with $i \in \{1,\dots,9\}$ at $t_2=46.9$ is 
presented. Note that the latter is again a point of maximal separation. Our analysis for the NOs implies the 
following intuitive picture for the beyond HF fermionic orbitals. The energetically lowest initially 
occupied orbitals cannot be excited to unoccupied excited states by interactions with another fermionic 
species for low interaction strengths, in contrast to the energetically higher initially occupied orbitals being more
susceptible to such excitations.   

In conclusion, we have studied a binary FF ensemble and showed that beyond HF effects 
can be observed even in the weakly interacting regime. The behavior of the ensemble on the level of NSFs 
is similar to the aforementioned BF case where the first NSF resembles the mean-field dynamics 
while the higher order contributions give rise to the observed beyond mean-field effects. 
Additionally, the structure of the NOs reveals that only higher-lying initially occupied 
fermionic states can be excited due to the presence of inter-species interactions.


\section{Conclusions}\label{chap:conc}
In this work, we present the ML-MCTDHX wave-function propagation method, which is designed for efficient 
ab-initio simulations of arbitrary multi-component ultracold atomic ensembles out of equilibrium. Employing 
both the concept of a  variationally optimally moving many-body basis and a multi-layer ansatz for the 
wave-function expansion allows for adapting ML-MCTDHX to system-specific intra- and inter-species 
correlations, which greatly improves its efficiency. By virtue of the so-called reduced transition matrices, we 
can formulate the ML-MCTDHX EOM in a unified manner, treating bosonic and fermionic components formally on an 
equal footing. Moreover, using the reduced transition matrices as central building blocks for evaluating the 
EOM proves to be computationally beneficial.

We demonstrate the power of the ML-MCTDHX method by applying it to the collision dynamics of few-atom binary 
mixtures of Bose-Fermi and Fermi-Fermi type in one dimension. Beginning with a detailed analysis of the convergence behavior 
of the densities, natural orbitals, and natural species functions by successively enlarging the employed 
basis sets, we show how the intra- and inter-species correlations are described with increasing accuracy. 
The truncation on the species layer is found to affect dominantly the accuracy 
of the computation, while, however, reducing the required number of coefficients for a converged simulation 
significantly: For e.g.\ the Fermi-Fermi collision scenario, a wave-function expansion w.r.t.\ to all numberstates
$\nk[A]{n_A}_t\nk[B]{n_B}_t$ would involve $K_AK_B\approx25\cdot10^6$ configurations, whereas the species-layer truncation
reduces the corresponding number of coefficients by a factor of about $500$ to $M^2+M(K_A+K_B)\approx 5\cdot10^4$.
Thereafter, we reveal the impact of correlations on the collision dynamics induced by the interspecies 
interaction. For a colliding BF mixture, we in particular observe a correlation-induced transmitted fraction, 
which is highly suppressed in corresponding mean-field calculations. This process takes place only in the 
sub-dominant natural species functions leading to a mixed state for the individual species. For a colliding FF 
mixture, we similarly find that, even at relatively weak interactions, correlations lead to a partial reflection 
via the sub-dominant natural species function. On the single-particle level, we classify the natural orbitals as 
core, valence and excited orbitals, where the population of the latter is a further signature of the beyond HF 
physics observed.

In the future, ML-MCTDHX will give access to the quantum dynamics of a multitude of physical mixture 
systems such that correlation-induced processes can be analyzed in detail. Here, we have in particular induced 
interactions, dephasing, self-localization, entangled state generation, as well as inter-species energy and 
momentum transfer in mind with a particular focus on the question how the distinguishability and particle 
statistics affects the physical behavior. Finally, we would like to stress that in order to unravel the correlated 
processes, further development of analysis tools for many-body systems needs to be put forward such as the 
here employed natural-species-function analysis shown to separate mean-field and correlated behavior. 
 
\begin{acknowledgments}
The authors acknowledge inspiring discussions with H.-D. Meyer. This work has been financially supported 
by the excellence cluster 'The Hamburg Centre for Ultrafast Imaging - Structure, Dynamics and Control of 
Matter at the Atomic Scale' of the Deutsche Forschungsgemeinschaft. 
Financial support by the Deutsche Forschungsgemeinschaft (DFG) in the framework of the
SFB 925 ''Light induced dynamics and control of correlated quantum
systems'' is gratefully acknowledged by L.C., S.I.M. and P.S. 
\end{acknowledgments}


\appendix
\section{Reduced few-body transition matrices}\label{app_trans_mat}

The reduced one and two-body density matrices, the mean-field operators and
the Hamiltonian matrix are key ingredients in the ML-MCTDHX EOM \eqref{eq:eom1}, \eqref{eq:eom2} and 
\eqref{eq:eom3}.
The calculation of these quantities shares a common
step, which is to calculate the reduced one- and two-body transition matrices:
\begin{align}\label{eq:tranmat}
\psiSb{i_\sigma} \ad{k} \a{q} \psiSk{j_\sigma}&= \sum_{\vec n|N_\sigma-1}Q_{\vec n}(k,q)\;
\big(C^\sigma_{i_\sigma,\vec n+\hat k}\big)^* C^\sigma_{j_\sigma,\vec n+\hat q}\\\nonumber
\psiSb{i_\sigma} \ad{k} \ad{q} \a{q'} \a{k'} \psiSk{j_\sigma} &=\sum_{\vec n|N_\sigma-2}
P_{\vec n}(k,q)P_{\vec n}(k',q')\;\big(C^\sigma_{i_\sigma,\vec n+\hat k+\hat q}\big)^* C^\sigma_{j_\sigma,\vec 
n+\hat k'+\hat q'}.
\end{align}
Here, $\hat k$ refers to an occupation number vector with zero occupations except for the 
occupation of the $k$-th SPF being set to unity. The particle-exchange symmetry enters the calculation only 
via the 
occupation number dependent factors $Q_{\vec n}(k,q)$ and $P_{\vec n}(k,q)$, which read for bosonic species
\begin{align}\label{eq:bosonicpq}
Q_{\vec n}(k,q)&=\sqrt{(n_k+1)(n_q+1)},\\
P_{\vec n}(k,q)&=\sqrt{(n_k+1+\delta_{kq})(n_q+1)},
\end{align}
 and for fermionic species
\begin{align}\label{eq:fermionicpq}
Q_{\vec n}(k,q)&=(-1)^{d_{\vec n}(k,q)},\\
P_{\vec n}(k,q)&=(1-\delta_{k,q})(-1)^{d_{\vec n}(k,q)+\theta(k,q)},
\end{align}
where we have defined 
\begin{equation}
 d_{\vec n}(k,q)=\sum_{a=\min(k,q)+1}^{\max(k,q)-1}n_a
\end{equation}
and $\theta(k,q)=1$ if $k>q$ and zero otherwise.

\section{Hamiltonian matrix}\label{app_hamilt_mat}

In order to calculate the Hamiltonian matrix
$\bra{\psiSlabel[A]{i_A}\psiSlabel[B]{i_B}}\hat H\ket{
\psiSlabel[A]{i_A'}\psiSlabel[B]{i'_B}}$
in SBS representation, as needed for the top-layer EOM \eqref{eq:eom1},
one may use the second-quantization representation of $\hat H$. Bearing in mind that the SBSs consist only of 
number states in which SPFs are occupied \footnote{But no other states of the full single-particle Hilbert 
space $\mathfrak{h}^\sigma$.}, only the creation and annihilation operators of SPFs provide non-vanishing 
contributions. Consequently, we find $\bra{\psiSlabel[A]{i_A}\psiSlabel[B]{i_B}} \hat H\ket{
\psiSlabel[A]{i_A'}\psiSlabel[B]{i'_B}}$ to be the sum of the following terms
\begin{align}
 \bra{\psiSlabel[A]{i_A}\psiSlabel[B]{i_B}}\big(\hat H_\sigma+\hat V_\sigma\big)\ket{
\psiSlabel[A]{i_A'}\psiSlabel[B]{i'_B}}
&=\sum_{r,s=1}^{m_\sigma} \;[ h_\sigma]_{s}^r\;
\psiSb{i_\sigma} \ad{r} \a{s} \psiSk{i_\sigma'}\\\nonumber
&\phantom{=}+\frac{1}{2}\sum_{r,s,u,v=1}^{m_\sigma}
[ v_\sigma]_{uv}^{rs}\;
\psiSb{i_\sigma} \ad{r} \ad{s} \a{v} \a{u} \psiSk{i_\sigma'}
\end{align}
with $[ h_\sigma]_{s}^r=\phiPb{r}\hat h_\sigma\phiPk{s}$ as well as
$[ v_\sigma]_{uv}^{rs}=\bra{\phiSlabel{r}\phiSlabel{s}}\hat v_\sigma \ket{\phiSlabel{u}\phiSlabel{v}}$, and 
\begin{align}
 \bra{\psiSlabel[A]{i_A}\psiSlabel[B]{i_B}}\hat W_{AB}\ket{\psiSlabel[A]{i_A'}\psiSlabel[B]{i'_B}}
&=\sum_{r,s=1}^{m_A}\sum_{u,v=1}^{m_B}[ w_{AB}]_{sv}^{ru}\;
\psiSb[A]{i_A} \ad[A]{r} \a[A]{s} \psiSk[A]{i_A'}\;\psiSb[B]{i_B} \ad[B]{u} \a[B]{v} \psiSk[B]{i_B'}
\end{align}
with $[ w_{AB}]_{sv}^{ru}=\bra{\phiSlabel[A]{r}\phiSlabel[B]{u}}
\hat w_{AB}\ket{\phiSlabel[A]{s}\phiSlabel[B]{v}}$ and the 
transition matrices given in Appendix~\ref{app_trans_mat}.

\section{Mean-field operator matrices}\label{app_mfop}

The mean-field operator matrices on the species-layer $[\hat{W}_{\sigma| \bar{\sigma}}]^i_j = 
\psiSb[\bar{\sigma}]{i}\hat W_{AB}\psiSk[\bar{\sigma}]{j}$ 
[see Eq.~\eqref{eq:eom2}] can be expressed in terms of the 
reduced one-body transition matrix as well
\begin{align}\label{eq:mfopmat_spec}
 [\hat{W}_{A|B}]^i_j &= 
\sum_{r,s=1}^{m_A}\sum_{u,v=1}^{m_B}[ w_{AB}]_{sv}^{ru}\;
\psiSb[B]{i} \ad[B]{u} \a[B]{v} \psiSk[B]{j}\; \ad[A]{r} \a[A]{s},\\
 [\hat{W}_{B|A}]^i_j &= 
\sum_{r,s=1}^{m_A}\sum_{u,v=1}^{m_B}[ w_{AB}]_{sv}^{ru}\;
\psiSb[A]{i} \ad[A]{r} \a[A]{s} \psiSk[A]{j}\; \ad[B]{u} \a[B]{v}.
\end{align}
On the particle layer, the corresponding intra- and inter-species mean-field operator matrices read
\begin{align}\label{eq:mfopmat_part_intra}
[\hat v_\sigma]^q_l&=\int {\rm d}y\,\phiPsb[\sigma]{y}{q}\,
v_\sigma(\hat x_1^\sigma,y)\,\phiPsk[\sigma]{y}{l},
\\\label{eq:mfopmat_part_inter1}            
[\hat
w_{A|B}]^q_l&=\int {\rm d}y\,\phiPsb[B]{y}{q}\,
w_{AB}(\hat x_1^A,y)\,\phiPsk[B]{y}{l},\\\label{eq:mfopmat_part_inter2}            
[\hat
w_{B|A}]^q_l&=\int {\rm d}y\,\phiPsb[A]{y}{q}\,
w_{AB}(y,\hat x_1^B)\,\phiPsk[A]{y}{l}.
\end{align}

\section{Density matrices}\label{app_dmat}
 The reduced density matrices, entering the EOM for the time-dependent basis states \eqref{eq:eom2} and
\eqref{eq:eom3}, can be calculated as follows. The reduced density matrix of the $\sigma$ species in SBS 
representation is given by
\begin{equation}\label{eq:eta1}
[\eta_{1,\sigma}]^i_j=\sum_{J^\sigma}\big(A_{J^\sigma_i}\big)^*A_{J^\sigma_j}
\end{equation}
 where the sum over the multi-index $J^\sigma$ denotes tracing out all the species but $\sigma$ and the
multi-index $J^\sigma_i$ refers to the corresponding configuration of all species with 
the index for the $\sigma$ species
being set to $i$. 
For a binary mixture, the reduced two-species density matrix reads $[\eta_{2,AB}]^{i_Ai_B}_{j_Aj_B}=
[\eta_{2,BA}]^{i_Bi_A}_{j_Bj_A}={\big(A_{i_Ai_B}\big)^*A_{j_Aj_B}}$, while for $S>2$ species we have
\begin{equation}\label{eq:eta2_intra}
 [\eta_{2,\sigma\kappa}]^{i_\sigma i_{\kappa}}_{j_\sigma 
j_{\kappa}}=\sum_{J^{\sigma\kappa}}\big(A_{J^{\sigma\;\kappa}_{i_\sigma 
i_{\kappa}}}\big)^*A_{J^{\sigma\;\kappa}_{j_\sigma j_{\kappa}}}
\end{equation}
 with the sum running over all SBSs of species other than $\sigma$, $\kappa$ and 
$J^{\sigma\;\kappa}_{i_\sigma 
i_{\kappa}}$ denoting the 
corresponding multi-index with the $\sigma$ ($\kappa$) species being in the $i_\sigma$-th ($i_\kappa$-th) SBS.

The reduced one-body density matrix of the $\sigma$ species can now be expressed as
\begin{equation}\label{eq:rho1}
[\rho_{1,\sigma}]^k_q=\sum_{i,j=1}^{M_\sigma}[\eta_{1,\sigma}]^i_j\psiSb{i} \ad{k} \a{q} \psiSk{j}.
\end{equation}
 Correspondingly, we find for the reduced two-body intra-species density matrix
\begin{align}\label{eq:rho2}
[\rho_{2,\sigma}]^{kq}_{uv}=&\sum_{i,j=1}^{M_\sigma}[\eta_{1,\sigma}]^i_j
\psiSb{i} \ad{k} \ad{q} \a{v} \a{u} \psiSk{j}
\end{align}
 Finally, the reduced two-body inter-species density matrix can be calculated as follows
\begin{align}\label{eq:rho2_inter}
[\rho_{2,\sigma\kappa}]^{kq}_{uv}&=\sum_{i_\sigma,j_\sigma=1}^{M_\sigma}\sum_{i_\kappa,j_\kappa=1}^{M_\kappa} 
[\eta_{2,\sigma\kappa}]^{i_\sigma i_{\kappa}}_{j_\sigma j_{\kappa}}\;
\psiSb[\sigma]{i_\sigma} \ad[\sigma]{k} \a[\sigma]{u} \psiSk[\sigma]{j_\sigma}
\psiSb[\kappa]{i_\kappa} \ad[\kappa]{q} \a[\kappa]{v} \psiSk[\kappa]{j_\kappa}.
\end{align}


\bibliography{FullBib}

\end{document}